\documentclass[aps,prl,floatfix,nofootinbib,superscriptaddress,twocolumn]{revtex4-1}
\usepackage{graphicx}  % needed for figures
\usepackage{dcolumn}   % needed for some tables
\usepackage{bm}        % for math
\usepackage{amssymb}   % for math
\usepackage{amsmath}
\usepackage{comment}
\usepackage{graphicx}
\usepackage{color}
\usepackage{slashed}
\usepackage{hyperref}
\usepackage[dvipsnames]{xcolor}

\usepackage{xr}

\begin{document}

\author{S\"{o}ren Schlichting}
\email{sschlichting@physik.uni-bielefeld.de}
\affiliation{Fakult\"{a}t f\"{u}r Physik, Universit\"{a}t Bielefeld, D-33615 Bielefeld, Germany}
\author{Sayantan Sharma}
\email{sayantans@imsc.res.in}
\affiliation{The Institute of Mathematical Sciences, a CI of Homi Bhabha National Institute, 
Chennai 600113, India}

\title{Chiral instabilities \& the fate of chirality imbalance in non-Abelian plasmas}
\date{\today}
\begin{abstract} 
We present a microscopic study of chiral plasma instabilities and axial charge transfer in 
non-Abelian plasmas with a strong gauge-matter coupling $g^2N_f=64$, by performing $3+1$ D real-time
classical-statistical lattice simulation with dynamical fermions. We explicitly demonstrate for the 
first time that -- unlike in an Abelian plasma -- the transfer of chirality from the matter sector 
to the gauge fields occurs predominantly due to topological sphaleron transitions. We elaborate on 
the similiarities and differences of the axial charge dynamics in cold Abelian $U(1)$ and non-Abelian 
$SU(2)$ plasmas, and comment on the implications of our findings for the study of anomalous transport 
phenomena, such as the chiral magnetic effect in QCD matter.

% At late stages of the evolution of this dynamical SU(2) 
% plasma, we thus do not observe any coherent long-distance magnetic fields  
% produced, instead the localized topological fluctuations survive. The 
% implication of our study for the chiral magnetic effect in QCD is discussed. 
\end{abstract}
\maketitle

\textit{Introduction.} 
Novel transport phenomena in the presence of chiral fermions have recieved a considerable amount 
of attention with manifestations from high-energy physics to condensed matter 
systems~\cite{Vilenkin:1980fu,Kharzeev:2013ffa,Miransky:2015ava,Kharzeev:2022ydx}. Specifically, 
the emergence of the Chiral Magnetic Effect (CME) 
~\cite{Kharzeev:2007jp,Fukushima:2008xe,Kharzeev:2012ph,Kharzeev:2015znc,
Koch:2016pzl} has been widely investigated in recent years, as a possible way 
to provide insights into the dynamics of topological structures in Quantum Chromodynamics
(QCD)~\cite{Kharzeev:2020jxw} or to realize transport phenomena in Dirac or Weyl 
semi-metals~\cite{Son:2012bg,Gorbar:2013dha,Li:2014bha,Cortijo:2016wnf,Kaushik:2018tjj,Sukhachov:2021fkh}.

Generically, such anomalous transport phenomena rely on the presence of a net chirality or axial charge 
imbalance $j^{0}_{5} \neq 0$ in the fermion sector. However, due to quantum effects, the axial current 
$j^{\mu}_{5}=(j^{0}_{5},\vec{\jmath}_{5})$ is not conserved, as expressed by the anomaly 
relation~\cite{Adler:1969gk, Bell:1969ts,Fujikawa:1980eg}
\begin{eqnarray}
\label{eq:AnomalyGen}
\partial_{\mu}j^{\mu}_{5}=-\frac{g^2N_{f}}{2\pi^2}\text{tr}[\mathbf{E\cdot B}]
\end{eqnarray}
valid for $N_f$ degenerate flavors of massless Dirac fermions in the presence of dynamical gauge fields. 
Notably the expression $\frac{g^2}{4\pi^2}\text{tr}[\mathbf{E\cdot B}]$ on the right hand side 
can be expressed as the divergence $\partial_{\mu}K^{\mu}$ of the Chern-Simons current $K^{\mu}$ 
describing the net helicity of the gauge fields, such that Eq.~(\ref{eq:AnomalyGen}) effectively 
describes the conservation of the combined net chirality of fermions and helicity of gauge fields. 
Since by virtue of Eq.~(\ref{eq:AnomalyGen}) the axial charge density $j^{0}_{5}$ in the fermion 
sector is manifestly not conserved, a crucial aspect of anomalous transport is therefore to understand 
how exactly and on what time scale the net chirality is transferred between fermionic and the gauge 
degrees of freedom.

Specifically, in the context of Abelian gauge theories such as Quantum Electrodynamcis (QED), it is well 
established~\cite{Hirono:2015rla,Gorbar:2016qfh,Tuchin:2017vwb,Mace:2019cqo} that fluctuations of the
electric and magnetic fields can deplete the net axial charge density in the system, thus requiring 
the application of external electromagnetic (EM) fields to sustain anomalous 
transport phenomena~\cite{Joyce:1997uy} such as the CME in condensed matter systems~\cite{Gorbar:2017lnp}. 
Conversely, in electromagnetic plasmas where no external fields are applied, any net chirality imbalance in 
the fermion sector is eventually transferred to the gauge field sector. Based on a series of studies based 
on weak-coupling techniques~\cite{Son:2012wh, Manuel:2015zpa,Gorbar:2016qfh,Tuchin:2017vwb}, 
effective macroscopic descriptions~\cite{Hirono:2015rla,Hattori:2017usa,Gorbar:2017vph} as 
well as non-perturbative real-time lattice 
simulation~\cite{Buividovich:2015jfa,Mace:2019cqo,Figueroa:2019jsi}, it has been established 
that the chirality transfer proceeds via an exponential growth-(and decay-) of right (left) 
handed helical magnetic field modes due to the so called chiral plasma instability~\cite{Akamatsu:2013pjd}.
Eventually, the unstable growth saturates, leading to a self-similar turbulent cascade that results in the 
generation of large scale helical magnetic fields~\cite{Mace:2019cqo,Hirono:2015rla,Rogachevskii:2017uyc}.

Despite the fact that different theoretical approaches are able to describe the dynamics of chirality 
transfer in Abelian gauge theories, the theoretical description of chirality transfer processes in 
non-Abelian gauge theories, such as QCD or the electroweak sector of the standard model, is significantly 
more involved. Due to the non-trivial structure of the gauge group, non-Abelian $SU(N_{c})$ gauge theories 
possess an infinite number of physically equivalent configurations whose vacua are topologically 
distinct~\cite{Mace:2016svc,Lenz:2001me}, that differ from each other by an integer amount of the chiral 
charge $N_{\rm CS}=\int_{\bf{x}} K^{0}$. Non-perturbative real-time processes can lead to so called 
sphaleron transitions between the different topological sectors \cite{Klinkhamer:1984di,Dashen:1974ck}, 
which tend to erase any pre-existing chiral charge imbalance in the fermion 
sector~\cite{McLerran:1990de,Cline:1993vv}, such that ultimately one expects all chiral charge to be 
absorbed into the topology of the non-Abelian gauge fields. Despite the fact that such processes are not 
only important for understanding the real-time dynamics of anomalous transport phenomena, but also play a 
crucial role in different scenarios of Baryo- or Leptogenesis in the early 
universe~\cite{Kuzmin:1985mm,Rubakov:1996vz,Bodeker:2020ghk}, a microscopic theoretical description is 
complicated by the interplay of several scales~\cite{Bodeker:1998hm}, and one typically resorts to 
effective macroscopic descriptions in the form of hydrodynamic equations~\cite{Erdmenger:2008rm, Son:2009tf, Loganayagam:2011mu, Kharzeev:2011ds,Banerjee:2012iz,Hongo:2013cqa} or classical effective 
theories~\cite{Akamatsu:2014yza,Akamatsu:2015kau,Mueller:2017arw}.

In this letter, we report on the first microscopic study of chirality transfer in non-Abelian plasmas. 
Specifically, we investigate the dynamics of chiral plasma instabilities and chirality transfer in an 
environment where the ambient temperature is much lower than the initial helicity chemical potential. 
Unlike at weak coupling and high temperatures where the primary unstable modes reside at
magnetic scale, there is no natural separation of scales 
in this regime. Starting from a large helicity imbalance in the fermion sector, we employ a 
classical-statistical description~ 
\cite{Polkovnikov:2009ys,Berges:2012ev,Kurkela:2012hp,Jeon:2013zga,Berges:2015kfa} to simulate 
the subsequent non-equilibrium evolution of the system. Beginning with the 
early onset of exponentially growing primary instabilities, we follow the evolution of the system 
to the highly non-linear regime where the unstable growth saturates and chirality transfer from the 
fermion to the gauge sector proceeds as anticipated via a sequence of correlated sphaleron transitions. 
We explicitly demonstrate that chirality transfer in non-Abelian plasmas is primarily driven by such
topological transitions and comment on the consequences of our findings for the emergence of anomalous
transport phenomena such as the CME in non-Abelian plasmas.

\textit{Simulation setup.} 
We perform real-time simulation of $N_f$ degenerate flavors of Dirac 
fermions of mass $m$, coupled to classical-statistical  
non-Abelian $SU(N_c)$ gauge fields~
\cite{Aarts:1998td,Berges:2010zv,Kasper:2014uaa,Mueller:2016aao,Muller:2016jod,Mace:2016shq}. 
We numerically solve the coupled set of Dirac equations 
\begin{align}
\label{eq:Dirac}
i\partial_t \hat{\Psi}_{\mathbf{x}}(t)&=\gamma^{0}(-i \gamma^i D_i[A]+m)\hat{\Psi}_{\mathbf{x}}(t)\,,
\end{align}
for the fermion fields $\hat{ \Psi}_\mathbf{x}(t)$,
where $D^i[A]\equiv \partial^i - ig A^i_\mathbf{x}(t)$ is 
the covariant derivative in temporal-axial ($A_0=0$) gauge, 
and the Yang-Mills' equations for the non-Abelian gauge fields 
$\mathbf{E}^a_\mathbf{x}(t)$ and $\mathbf{B}^a_\mathbf{x}(t)$,
\begin{align}
\label{eq:Maxwell}
\partial_t g\mathbf{E}^a_\mathbf{x}(t) &- [\mathbf{D} \times g\mathbf{B}_\mathbf{x}(t)]^a = -g^2 N_{f} \, \mathbf{j}_{\mathbf{x}}^a(t)\,.
\end{align}
Evolution equations (\ref{eq:Dirac}) and (\ref{eq:Maxwell}) include the effects of color 
fields on the fermion sector as well as the non-linear back-reaction of fermion currents 
$\mathbf{j}_{\mathbf{x}}^{a}(t)=\langle \frac{1}{2} 
[ \hat{{\Psi}}^\dagger_\mathbf{x}(t) t^{a}\gamma^0\bm{\gamma} ,\hat{ \Psi}_\mathbf{x}(t)]\rangle$,
on the dynamical evolution on the non-Abelian fields in Eq.~(\ref{eq:Maxwell}). We note that
the classical-statistical description in Eqns.~(\ref{eq:Dirac},\ref{eq:Maxwell}) 
is accurate to leading order in the gauge coupling $g^2$, but to all orders 
in the coupling $g^2N_f$ between matter and gauge fields~\cite{Aarts:1998td,Kasper:2014uaa}. 
In this work we focus on the simplest non-Abelian gauge group $SU(2)$ and employ $g^2N_f=64$ 
to implement a strong back-reaction of the matter fields to the gauge fields. This allows us 
not only to properly resolve all relevant scales of the problem, but also to investigate the 
mechanism of the chirality transfer beyond the weak-coupling limit~\cite{Akamatsu:2013pjd,Akamatsu:2015kau}.

We formulate the problem on a $N_s^{3}$ spatial lattice with
lattice spacing $a_s$, using a compact Hamiltonian lattice formulation
of $SU(2)$ gauge theory~\cite{Kogut:1974ag}, with $O(a_s^3)$ tree-level 
improved Wilson-fermions~\cite{Mace:2016shq} which is essential for 
realizing the chiral anomaly with a good precision on a finite size
lattice. Clearly, the implementation of the fermion field operator 
$\hat{\Psi}_{\mathbf{x}}(t)$ is the most expensive part of our 
numerical algorithm as the solution to the operator Eq.  
(\ref{eq:Dirac}) is constructed from linear combinations of a 
complete set of $8 N_s^3$ wave-functions~\cite{Aarts:1998td,Mace:2016shq}.

We will analyze the non-equilibrium evolution of a chirally imbalanced charge 
neutral ensemble of fermions, by specifying the initial occupation numbers 
of left~(L) and right~(R) handed fermions at $t=0$ according to a 
Fermi-Dirac distribution $n_F^{L/R}(t=0,\mathbf{p})=\frac{1}{e^{(E_{\mathbf{p}} \pm 
\mu_{h})/T}+1}$ with a helicity chemical potential $\mu_h$ and energy 
$E_{\mathbf{p}}=\pm \sqrt{\mathbf{p}^2+m^2}$ for particles and 
anti-particles respectively. The initial conditions for the gauge 
fields are chosen as a classical-statistical ensemble representing 
vacuum fluctuations~\cite{Aarts:2001yn,Mace:2019cqo}. We focus on 
the dynamics in a cold and dense plasma with an initial $T/ \mu_h = 1/8$ 
to make a comparison with our earlier studies of an Abelian plasma~\cite{Mace:2019cqo} 
and please refer to the supplemental material of~\cite{Mace:2019cqo} for 
additional details of the implementation.
% A leap-frog solver with time step $a_t=0.001 a_s$ is implemented to 
% solve the discretized equations of motions.
To estimate the residual effects of finite lattice spacing and lattice size, 
we have performed variations of $\mu_{h} a_s=0.8$ -$1.0$ and $N_s=24$ -$40$
and if not stated otherwise, we will present results for our largest and finest 
lattices with $N_s=40$ and $\mu_{h}a_s=0.8$. Simulations are performed close to 
the chiral limit $m \ll \mu_h$ by fixing $m a_s=5\cdot 10^{-4}$. We will express 
all results in the units of the only dimensionful scale, the helicity chemical $\mu_h$. 
By appropriate choice of $\mu_{h}$, this allows for a translation of the physical time 
scales in different physical settings.  So for instance if in a cosmological setting 
$\mu_{h}=100 ~{\rm GeV}$ and $T \ll \mu_{h}$ then the time scale $1/\mu_{h} \approx 0.002 ~{\rm fm}/c$. 
Generally, changing from $SU(2)$ to $SU(N)$ can be expected to change the order one pre-factors 
in the dynamics. Similarly, the dependence on the gauge coupling $g$ is only logarithmic, 
as is typically the case for the dynamics of instabilities, so the only relevant parameter 
affecting the time scales is the gauge matter coupling $g^2N_f$. While it would be interesting 
to investigate the dependence on $g^2N_f$ further, the numerical cost of the simulations 
prohibits a full exploration of the parameter space.

%All the results discussed below will be expressed in units of dimensionless quantities in units of the initial helicity chemical potential $\mu_{h}$. 

\begin{figure}[th]
\begin{center}
\includegraphics[width=0.45\textwidth]{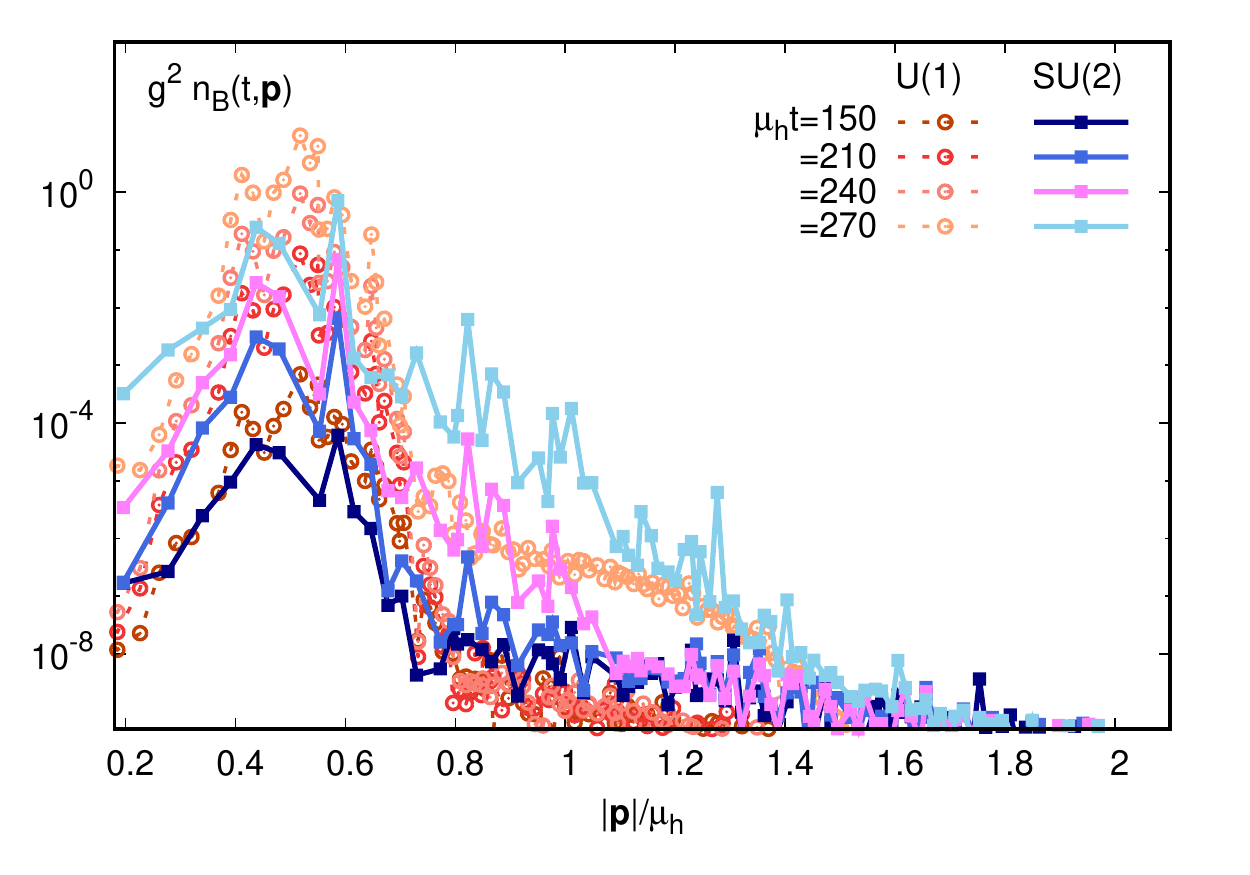}
\end{center}
\caption{Evolution of the (chromo-) magnetic field spectra $g^2n_{B}(t,{\bf p})$ at early 
times $\mu_h.t< 300$ in an $SU(2)$ plasma ($V=32/\mu_h^3~,\mu_h a_s=0.8$) shown 
in shades of blue and a $U(1)$ plasma ($V=48/\mu_h^3~,\mu_h a_s=1$), shown in 
shades of orange. Chiral plasma instabilities lead to the exponential growth of 
right-handed magnetic field modes with momenta $|{\bf p}|/\mu_{h} \sim 0.5$ with 
comparable growth rates in $SU(2)$ and $U(1)$ plasmas.}
\label{fig:GaugeSpectraSU2vsU1Early}
\end{figure}

\textit{Chiral plasma instabilities.} 
We now proceed to analyze the mechanism of chirality transfer from 
fermions to gauge fields, and frequently contrast our findings for non-Abelian plasmas 
with earlier studies in Abelian gauge theories~\cite{Mace:2019cqo}. Starting from a large 
net chirality imbalance in the fermion sector, both Abelian and non-Abelian plasmas exhibit 
an instability~\cite{Akamatsu:2013pjd} resulting in an exponential growth/suppression of 
right/left handed magnetic field modes at early times. We illustrate this behavior in 
Fig.~\ref{fig:GaugeSpectraSU2vsU1Early}, where we present the evolution of the spectrum 
of (chromo-)magnetic field modes
\begin{align}
\label{eq:nBDef}
n_{B}(t,\mathbf{p})=\frac{1}{\nu}\sum_{a=1}^{N_c} \frac{| \mathbf{B}^a(t,\mathbf{p}) |^2}{|\mathbf{p}|}
\end{align}
where $\mathbf{B}^a(t,\mathbf{p})=\frac{1}{\sqrt V}\int d^3\mathbf{x}~\mathbf{B}^a(\mathbf x,t)
\rm{e}^{-i\mathbf{p\cdot x}}$ is the Fourier transform of the (chromo-)magnetic field 
strength $\mathbf{B}^a(\mathbf x,t)$ extracted from elementary lattice plaquettes as in~\cite{Mace:2019cqo} 
and $\nu=(N_c^2-1)$ for $SU(N_c)$ and $\nu=1$ for the $U(1)$ gauge group respectively.
\footnote{Since the magnetic field strength $B^a(\mathbf x,t)$ transforms non-trivially 
under non-Abelian gauge transformations, we follow ~\cite{Berges:2013fga,Kurkela:2012hp} 
and calculate the equal time correlation function in Eq.~(\ref{eq:nBDef}) in Coulomb gauge 
to minimize gauge artifacts.} 

Starting from initial vacuum fluctuations, the chiral plasma instability results in a rapidly 
growing population of primarily unstable modes with $|\mathbf{p}|/\mu_h \lesssim 0.8$. Small 
momentum modes with $|\mathbf{p}|/\mu_h\sim 0.5$ feature the largest growth rates, resulting 
in a pronounced peak in the spectra. Since at early times $\mu_h t \lesssim 300$ the occupation 
numbers are small, $g^2 n_{B}(t,\mathbf{p}) \ll 1$, such that the field strength remains 
perturbative, the qualitative behavior of the Abelian $U(1)$ and non-Abelian $SU(2)$ theories 
is essentially the same; the only notable exception is the earlier onset of secondary instabilities 
for the latter due to non-Abelian self-interactions~\cite{Berges:2007re} around $\mu_h t \sim 200$, 
which results in the population of higher momentum modes $|\mathbf{p}|/\mu_h \gtrsim 1$ for the 
$SU(2)$ plasma.

\begin{figure}[tb]
\begin{center}
\includegraphics[width=0.45\textwidth]{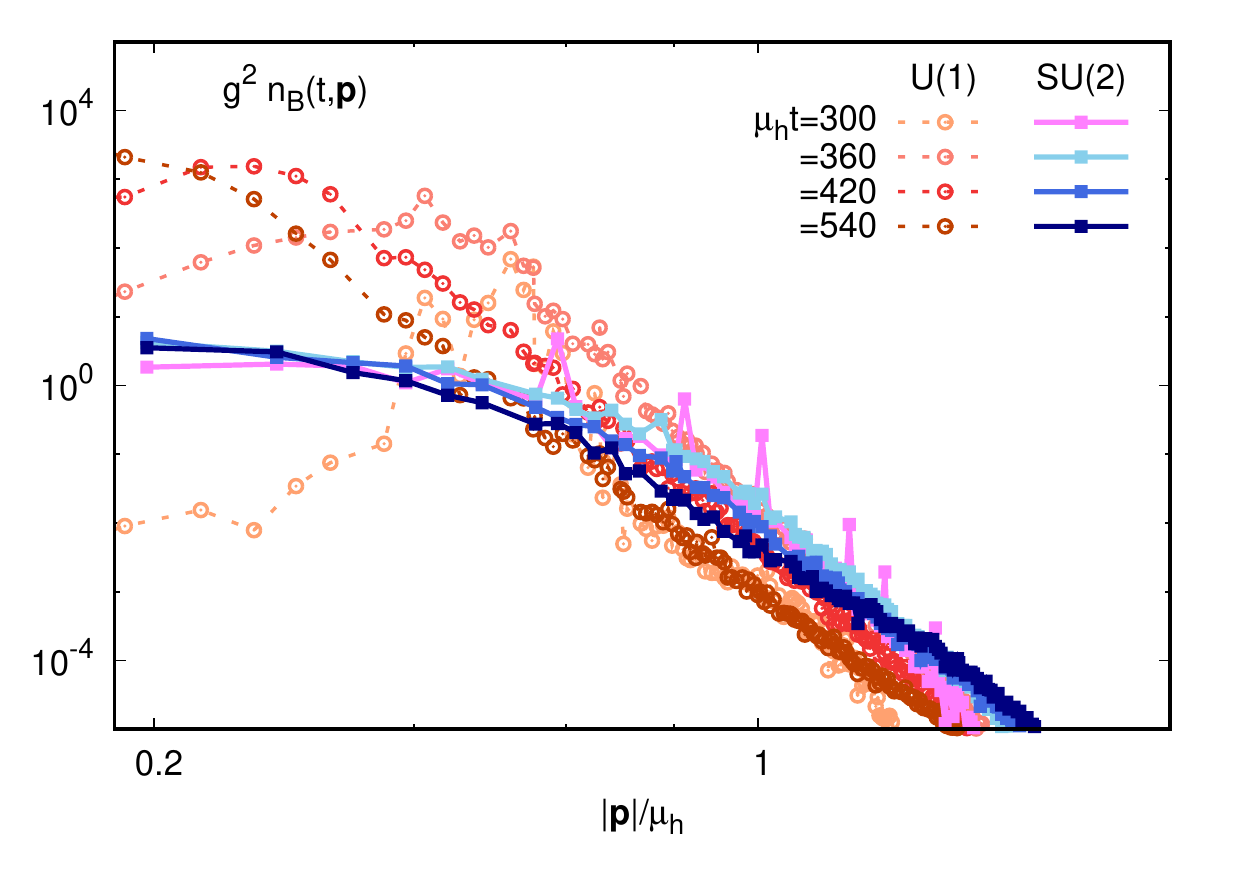}
\end{center}
\caption{Evolution of the (chromo-) magnetic field spectra $g^2n_{B}(t,|{\bf p}|)$ at 
late times $\mu_h.t> 300$ in an $SU(2)$ plasma ($V=32/\mu_h^3~,\mu_h a_s=0.8$) shown 
in shades of blue and a $U(1)$ plasma ($V=48/\mu_h^3~,\mu_h a_s=1$), shown in 
shades of orange. Due to non-Abelian self-interactions the unstable growth saturates 
for the $SU(2)$ plasma and the magnetic field occupation numbers do not exceed the
non-perturbative threshold $n_{B}(t,|{\bf p}|) \sim 1/g^2$. Conversely, the Abelian 
$U(1)$ plasma shows an inverse cascade of magnetic helicity, resulting in the generation 
of strong large scale coherent magnetic fields~\cite{Mace:2019cqo}. }
\label{fig:GaugeSpectraSU2vsU1Late}
\end{figure}

Striking differences between the evolution in the Abelian and the non-Abelian plasma start to emerge 
for times $\mu_h t \gtrsim 300$ depicted in Fig.~\ref{fig:GaugeSpectraSU2vsU1Late}, where the occupation 
numbers become non-perturbatively large, $g^2 n_{B} \sim 1$, and the dynamics becomes highly non-linear. 
Specifically, for the $SU(2)$ gauge theory, the non-Abelian self-interactions of the gauge fields lead to 
saturation of the unstable growth once the occupation numbers $n_{B}$ of unstable modes become on the 
order of the inverse self-coupling $\sim 1/g^2$~\cite{Berges:2007re,Kurkela:2011ti}. While at intermediate 
times $\mu_{h} t \sim 300$ interactions between unstable modes produce distinct peaks in the spectrum at 
integer multiples of the momentum of the primarily unstable mode $|\mathbf{p}|/\mu_{h} \sim 0.5$ 
successive interactions lead to rapid population of the ultra-violet tails of the spectrum. Subsequently
for $\mu_{h} t \gtrsim 360$, the spectrum of the (chromo-)magnetic $SU(2)$ fields features a large 
infrared occupation $n_{B} \sim 1/g^2$ for $|\mathbf{p}|/{\mu}_{h} \lesssim 0.5$ followed by a rapid 
decrease towards the ultra-violet and approximately retains this shape over the course of the entire 
evolution depicted in Fig.~\ref{fig:GaugeSpectraSU2vsU1Late}. 

Conversely, in the case of the Abelian $U(1)$ gauge theory, the  growth of unstable modes only saturates 
at a later time leaving a pronounced peak in the spectrum around $|\mathbf{p}|/\mu_{h} \sim 0.5$ for 
$\mu_{h} t \sim 360$. Subsequently, the characteristic peak moves towards lower and lower momenta, where 
in sharp contrast to the $SU(2)$ plasma, infrared occupation numbers at late times $\mu_h t \gtrsim 360$ 
do exceed the non-perturbative threshold $n_{B} \gtrsim 1/g^2$. Strikingly, this behavior can be 
associated with a self-similar inverse cascade of the magnetic helicity, where the evolution of the
spectrum can (approximately) be described in terms of universal scaling functions and scaling 
exponents~\cite{Mace:2019cqo}. Over the course of this process, the net axial charge of the fermions is 
transferred to the gauge field sector, and subsequently transported to lower and lower momentum scales, 
eventually resulting in the generation of long range helical magnetic fields~\cite{Mace:2019cqo,Hirono:2015rla,Rogachevskii:2017uyc}.

\textit{Chirality transfer \& fate of axial charge imbalance in non-Abelian plasmas.} %Understanding the helicity transfer mechanism in $SU(2)$
Since in contrast to the Abelian $U(1)$ plasma no long range coherent fields are generated 
in the case of a non-Abelian $SU(2)$ plasma (see also ~\cite{Kurkela:2011ti,Kurkela:2012hp}) 
it becomes a crucial question to what extent and by which mechanism axial charge is transferred 
from fermions to gauge fields over the course of the instability dynamics. In order to investigate 
the chirality transfer mechanism in non-Abelian plasmas, it proves insightful to study the evolution 
of the Chern-Simons number 
\begin{align}
\label{eq:NCs}
\Delta N_{CS}(t)
&=\frac{g^2}{4\pi^2}\int_{0}^{t}~dt^{'}\int_{{\bf x}} \text{Tr}\left[ \mathbf{E}_{\bf x}(t^{'})\cdot\mathbf{B}_{\bf x}(t^{'})\right]\;,
\end{align}
which represents the gauge field contribution to the anomaly budget in Eq.~(\ref{eq:AnomalyGen}), 
%(i.e. $\Delta N_{CS}(t)=\int_{0}^{t}dt'\int_{{\bf x}} \partial_{\mu}K^{\mu}$)
such that when integrated over space the balance equation (\ref{eq:AnomalyGen}) for the axial charge 
takes the form~\footnote{Note that strictly speaking this balance equation is only valid for $N_f$ 
flavors of massless fermions. However, since in practice the fermion mass $m/\mu_h\ll 1$ is sufficiently
small, finite mass effects are negligible over the time scales of our simulations and we have verified 
this explicitly by tracking the corresponding contribution to the anomaly budget.} 
\begin{align}
\frac{\Delta J_{5}^{0}(t)}{N_f}= -2\Delta N_{\rm CS}(t)\;.
\end{align}

\begin{figure}[t!]
\begin{center}
\includegraphics[width=0.5\textwidth]{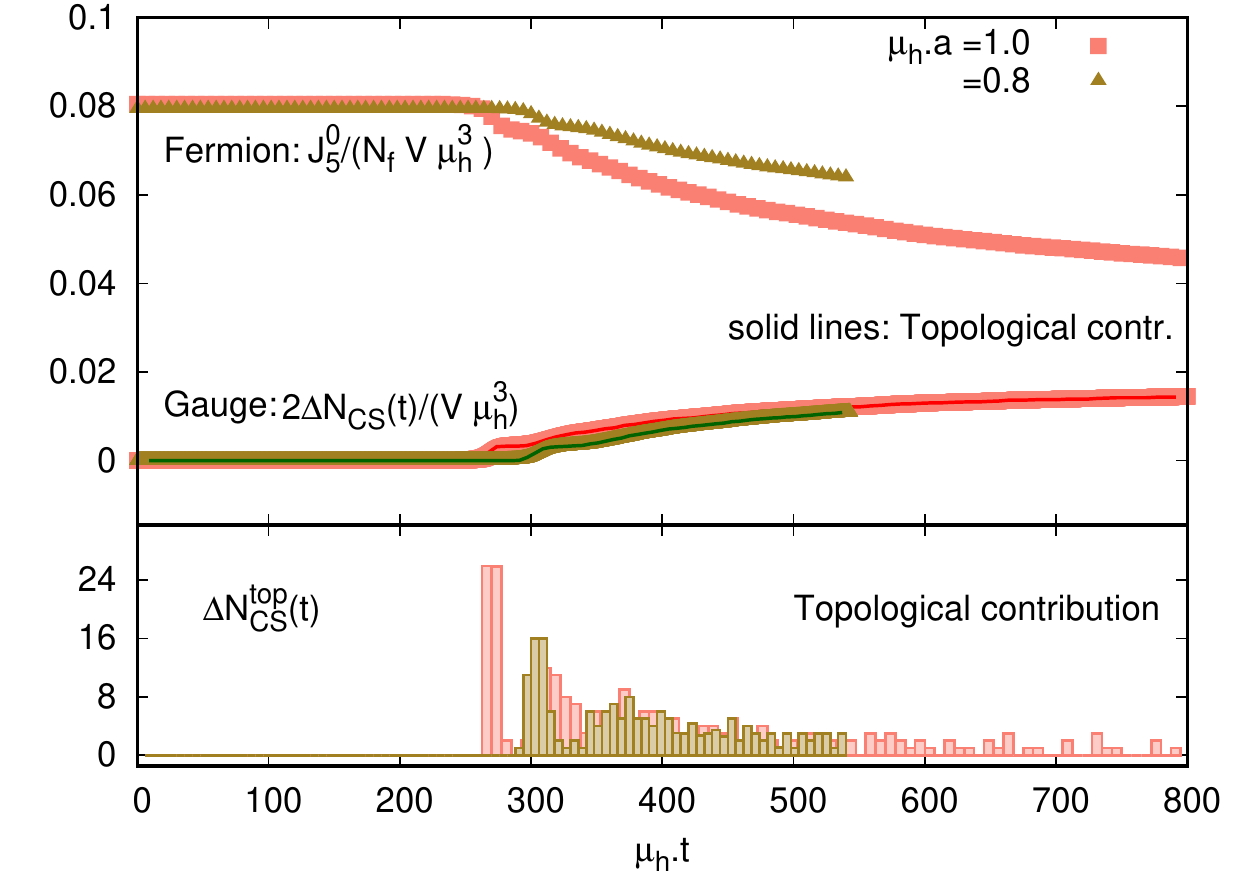}
\end{center}
\caption{(top) Evolution of the axial charge density of fermions $J_{5}^{0}/(N_f \mu_{h}^3 V)$ and gauge fields 
$2\Delta N_{\rm CS}/(\mu_{h}^3 V)$ in the $SU(2)$ plasma. Different symbols show the results for two different 
lattice spacings $\mu_h a_s=0.8$ (green) and $\mu_h a_s=1$ (red) respectively. Changes in the Chern-Simons number 
$\Delta N_{\rm CS}$ are predominantly due to non-equilibrium sphaleron transitions, as shown by the solid line, which 
contains the topological contribution obtained by cooling the non-Abelian gauge field configurations. (bottom) Change 
in the Chern-Simons number $\Delta N_{\rm CS}^{\rm top}(t)$ over a short time interval $\Delta t/a_s=7.5$ due to 
topological transitions. }
\label{fig:AnomalyBudget}
\end{figure}

Our results for the chirality transfer in the non-Abelian $SU(2)$ plasma are depicted in the top panel of 
Fig.~\ref{fig:AnomalyBudget}, where we present the time evolution of the axial charge densities of fermions
[$J_5^{0}/(VN_f)$] and net-helicity of gauge fields $(2 \Delta N_{\rm CS}(t)/V)$. Different curves show the
results for two different lattice discretizations ($N_s=40, ~\mu_h a_s=0.8$ and $N=32, \mu_h a_s=1$),
indicating excellent convergence for $\Delta N_{\rm CS}$, whereas a sufficiently fine discretization 
is necessary to properly resolve the evolution of the axial charge of fermions $J_5^{0}$ and satisfy 
the anomaly relation in the lattice discretized theory (also 
see~\cite{Mace:2016shq,Karsten:1980wd}).\footnote{We find that for $\mu_{h}a_s=0.8$ residual violations of 
the anomaly relation are always below the $\sim 5\%$ level even at late times $\mu_h t \gtrsim 500$.}

Starting around $\mu_{h} t\approx 300$ where the chiral plasma instability saturates, one observes an
initially rapid transfer of chirality from fermions to gauge fields which subsequently slows down and
continues over the course of the entire evolution in Fig.~\ref{fig:AnomalyBudget}. While at first sight, 
the evolution of $J_5^{0}/(VN_f)$ and $2 \Delta N_{\rm CS}(t)/V$, normalized in units of $\mu_h^3$ in 
the non-Abelian plasma appears to be similar to the corresponding results in the Abelian $U(1)$ case
presented in~\cite{Mace:2019cqo}, there is a crucial difference with regards to the nature of the
configurations of the gauge fields that carry the net axial charge. Specifically it turns out that -- 
in sharp contrast to Abelian gauge theories -- the changes in Chern-Simons number can be attributed
to the change in topology of the non-Abelian gauge field configurations emerging from a correlated sequence 
of sphaleron transitions. In order to isolate the topological contribution to $\Delta N_{\rm CS}$, we follow 
standard procedure and perform a gauge covariant cooling of the $SU(2)$ gauge field 
configurations~\cite{Moore:1998swa}, which removes all excitations carrying a finite amount of energy while 
leaving the topology of the configuration untouched (see Appendix for details). 

The bottom panel of Fig.~\ref{fig:AnomalyBudget} shows the change in Chern-Simons number 
$\Delta N_{\rm CS}^{\rm top}(t)=\Delta N_{\rm CS}^{\rm cooled}(t+\Delta t/2) - \Delta N_{\rm CS}^{\rm 
cooled}(t-\Delta t/2)$ due to topological transitions over a small time interval $\Delta t/a_s=7.5$, and can 
essentially be understood as the (discrete) time derivative of the solid curves in the top panel (times $V 
\Delta t$). Due to its topological nature, this quantity is integer valued and  represents the number of 
sphaleron transitions over the time interval $\Delta t$. In contrast to thermal ensembles where individual 
sphaleron transitions are observed to be uncorrelated and the Chern-Simons number $\Delta N_{\rm CS}(t)$ 
exhibits an integer random walk behavior (see e.g.~\cite{Moore:1998swa}), individual sphaleron transitions 
are highly correlated as all of them tend to erase the net axial charge in the fermion sector. 

By inspecting the solid lines in the top panel, which represent the time integrated contribution to the 
Chern-Simons number from topological transitions $\sum_{t'<t} \Delta N_{\rm CS}^{\rm top}(t')$, we find that 
topological sphaleron transitions essentially account for the entire change in the Chern-Simons number. We 
therefore conclude from this analysis that on sufficiently large time scales, the initial axial charge 
imbalance is completely erased and absorbed by the topology of the non-Abelian gauge fields.

\textit{Conclusions \& Outlook.} 
In this letter we report on the microscopic mechanism of the transfer of net 
chirality from the matter sector to the gauge field sector in a chirally
imbalanced non-Abelian plasma. Based on classical-statistical 
lattice simulations close to the continuum limit, we can establish that 
the chirality transfer in non-Abelian plasmas occurs predominantly via topological transitions. 
Specifically, for the cold plasmas investigated in this study, the process is initiated by the onset of 
chiral plasma instabilities in the gauge field sector, which
ultimately triggers very frequent sphaleron transitions, leading to a rapid transfer of chirality from 
the fermions to the gauge sector. Since the axial charge is absorbed by the topology of the gauge field, 
the gauge field occupancies do not exceed the scale $1/g^2$, and no coherent 
long-range magnetic fields are generated throughout the process. Strikingly, this is in sharp contrast to 
Abelian gauge theories, where as reported previously~\cite{Mace:2019cqo} a chirality imbalance in the 
fermion sector ultimately leads to the generation of large scale helical magnetic fields via an inverse 
turbulent cascade of the magnetic helicity. Conversely, in non-Abelian gauge theories the initial 
chirality imbalance is completely erased, and as the topologically distinct gauge field  configurations 
are physically indistinguishable there is no physical effect on the evolution of the plasma on large 
time scales.

Our study provides the first explicit microscopic demonstration of the washout of an initial 
chiral charge imbalance due to topological transitions in a cold $SU(2)$ plasma and has important 
implications for the study of anomalous transport phenomena which are driven by a net chirality 
imbalance in the fermion sector. Since the net chirality imbalance is ultimately erased, effects 
like the CME should be understood as transient non-equilibrium phenomena that persist only over a 
limited time scale. 

Since unlike in the present study the QGP created in heavy-ion collisions is initially in a hot 
state with $T \gg \mu_{h}$, thus frequent changes of the topology of the gauge fields occur, even 
in the absence of an axial charge imbalance due to thermal sphaleron 
transitions~\cite{McLerran:1990de}. In the presence of an axial charge imbalance in the fermion 
sector, sphaleron transitions exhibit a bias, which just like in the present study, tends to 
erase the axial charge imbalance in the fermion sector, such that an initial charge imbalance 
decays on a time scale $\tau_{\rm sph} = \chi_{A}T / \Gamma_{\rm sph}$ where $\chi_{A}$ 
is the axial charge susceptibility and $\Gamma_{\rm sph}$ is the sphaleron transition 
rate~\cite{McLerran:1990de}. Based on recent lattice estimates of $\Gamma_{\rm sph} \sim 0.1 
T^4$~\cite{BarrosoMancha:2022mbj,Altenkort:2020axj} and the axial charge susceptibility of the 
pure gauge theory $\chi_{A} = N_cT^2/3$, this time scale is of the order of $\sim 10/T$ and we 
therefore anticipate that in heavy-ion collisions local pockets of non-zero net chirality created 
during the early non-equilibrium stages will slowly disappear, while at the same time more local 
regions of net chirality imbalance will be created due to finite temperature sphaleron 
transitions.
    
Evidently, a more detailed microscopic study of these competing phenomena in a QCD plasma will 
require the extension of our study to hot plasma, which requires us to disentangle the dynamics 
of the infrared modes $|\mathbf{p}|\lesssim g^2T$ that are primarily responsible for chirality 
transfer from that of the (noisy) ultraviolet modes which reside at the scale $\sim 
T$~\cite{Akamatsu:2014yza,Akamatsu:2015kau}. Since the ultimate fate of a chiral charge 
imbalance is quite different in Abelian and non-Abelian gauge theories, it would also be 
interesting to study these competing effects in theories like the standard model of particle 
physics, where matter fields are simultaneously coupled to both Abelian and non-Abelian gauge 
fields.

\textit{Acknowledgements.}
  We thank N.~Mueller, M.~Mace and D.~B\"odeker for discussions and 
  collaboration on closely related projects. This work is supported 
  in part by the Deutsche Forschungsgemeinschaft 
  (DFG, German Research Foundation) through the CRC-TR 211 
  ’Strong-interaction matter under extreme conditions’– project 
  number 315477589 – TRR 211.
  Sa.S. gratefully acknowledges partial support from the Department of 
  Science and Technology, Govt. of India through a Ramanujan fellowship 
  and from the Institute of Mathematical Sciences. This research used 
  resources of the National Energy Research Scientific Computing Center 
  (NERSC), a U.S. Department of Energy Office of Science User Facility 
  operated under Contract No. DE-AC02-05CH11231. Simulations were also 
  performed on the national supercomputers Cray XC40 Hazel Hen and HPE 
  Apollo Hawk at the High Performance Computing Center Stuttgart (HLRS) 
  under the grant number 44156 (CMESIMULATION).

 \textit{{\bf Appendix:} Topological contribution to Cherns-Simons number.} Below we explain the extraction of the topological contribution to the change in the Chern-Simons number $\Delta N_{\rm CS}^{\text{top}}(t)$. We follow~\cite{Mace:2016svc} and employ gradient flow cooling of the three dimensional 
  gauge link configurations $U_{\mathbf{x},i}(t-\Delta t/2,\tau_c=0)$ and 
  $U_{\mathbf{x},i}(t+\Delta t/2,\tau_c=0)$ up to a cooling time $\tau_c=512 a_s^2$ by performing 
  $4096$ cooling steps with step size $\delta \tau_{c}/a_s^2=1/8$. Since each 
  cooling step reduces the average field strength, this effectively results in 
  three-dimensional pure gauge configurations
  %, with typical magnetic energy densities $\lesssim XXX$
  at the end of the cooling. Next, in order to compare 
  the topology of the cooled configurations at times $t-\Delta/2$ and $t+\Delta t/2$ 
  with $\Delta t/a_s=7.5$, we connect the two independently cooled gauge configurations
  $U_{\mathbf{x},i}(t-\Delta t/2,\tau_c=0)$ and $U_{\mathbf{x},i}(t+\Delta t/2,\tau_c=0)$ 
  via a geodesic interpolation on the $SU(2)$ group manifold, i.e. we express
\begin{eqnarray}
\label{eq:Econnection}
U_{\mathbf{x},i}(t+\frac{\Delta t}{2},\tau_c) = \rm{e}^{i g a_{s}^2 \mathcal{E}_{\mathbf{x}}^{a}t_{a} \Delta t} U_{\mathbf{x},i}\left(t-\frac{\Delta t}{2},\tau_c\right)
\end{eqnarray}
and construct interpolating links
\begin{eqnarray}
\label{eq:Ugeodesic}
U_{\mathbf{x},i}(t+\theta \Delta t,\tau_c)=
\rm{e}^{i g a_{s}^2 \mathcal{E}_{\mathbf{x}}^{a}t_{a} (\theta+\frac{1}{2})\Delta t}
U_{\mathbf{x},i}\left(t-\frac{\Delta t}{2},\tau_c\right)
\end{eqnarray}
for $-1/2< \theta < 1/2$ to calculate the difference in the Chern-Simons number as
\begin{eqnarray}
\label{eq:NCStopDef}
\Delta N_{\rm CS}^{\text{top}}(t)=N_{\rm CS}^{\rm cooled}\left(t-\frac{\Delta t}{2}\right)-N_{\rm CS}^{\rm 
cooled}\left(t+\frac{\Delta t}{2}\right) \\
=\frac{1}{8\pi^2} \frac{\Delta t}{a_s} \int_{-1/2}^{1/2}d\theta~ \sum_{\mathbf{x}} \left.  
g a_s^2\mathcal{E}_{\mathbf{x}}^{a}~ ga_S^2\mathcal{B}_{\mathbf{x}}^{a}(t+\theta\Delta t)
\right|_{\tau_c}\;, \nonumber
\end{eqnarray}
where $\mathcal{E}_{\mathbf{x}}^{a}$ denote the fields in Eq.~(\ref{eq:Econnection}), 
$\mathcal{B}_{\mathbf{x}}^{a}(t+\theta\Delta t)$ is the magnetic field strength associated 
with the  interpolating gauge links in Eq.~(\ref{eq:Ugeodesic}). We employ an $\mathcal{O}(a_s^2)$ 
improved discretization of $\mathcal{E}$ and $\mathcal{B}$~\cite{Moore:1996wn,Mace:2016svc}, 
and calculate the integral over $\theta$ using Simpsons rule with $N_{\theta}=16$ integration 
points. 

Since Eq.~(\ref{eq:NCStopDef}) provides the topological contribution to the change in Chern-Simons 
number $\Delta N_{\rm CS}$ over the time interval $[t-\Delta t/2,t+\Delta t/2]$ shown in the bottom 
panel of Fig.~\ref{fig:AnomalyBudget}, the topological contribution to the change in Chern-Simons 
number over the time interval $[0,t]$, shown in the top panel of Fig.~\ref{fig:AnomalyBudget}, can 
then be computed as the sum over the contributions from each interval, i.e.
\begin{eqnarray}
\left. \Delta N_{\rm CS}(t) \right|^{\rm topological}_{\rm contribution} = 
\sum_{t'<t} \Delta N_{\rm CS}^{\text{top}}(t')\;.
\end{eqnarray}

\bibliography{references.bib}

\begin{thebibliography}{73}%
\makeatletter
\providecommand \@ifxundefined [1]{%
 \@ifx{#1\undefined}
}%
\providecommand \@ifnum [1]{%
 \ifnum #1\expandafter \@firstoftwo
 \else \expandafter \@secondoftwo
 \fi
}%
\providecommand \@ifx [1]{%
 \ifx #1\expandafter \@firstoftwo
 \else \expandafter \@secondoftwo
 \fi
}%
\providecommand \natexlab [1]{#1}%
\providecommand \enquote  [1]{``#1''}%
\providecommand \bibnamefont  [1]{#1}%
\providecommand \bibfnamefont [1]{#1}%
\providecommand \citenamefont [1]{#1}%
\providecommand \href@noop [0]{\@secondoftwo}%
\providecommand \href [0]{\begingroup \@sanitize@url \@href}%
\providecommand \@href[1]{\@@startlink{#1}\@@href}%
\providecommand \@@href[1]{\endgroup#1\@@endlink}%
\providecommand \@sanitize@url [0]{\catcode `\\12\catcode `\$12\catcode
  `\&12\catcode `\#12\catcode `\^12\catcode `\_12\catcode `\%12\relax}%
\providecommand \@@startlink[1]{}%
\providecommand \@@endlink[0]{}%
\providecommand \url  [0]{\begingroup\@sanitize@url \@url }%
\providecommand \@url [1]{\endgroup\@href {#1}{\urlprefix }}%
\providecommand \urlprefix  [0]{URL }%
\providecommand \Eprint [0]{\href }%
\providecommand \doibase [0]{http://dx.doi.org/}%
\providecommand \selectlanguage [0]{\@gobble}%
\providecommand \bibinfo  [0]{\@secondoftwo}%
\providecommand \bibfield  [0]{\@secondoftwo}%
\providecommand \translation [1]{[#1]}%
\providecommand \BibitemOpen [0]{}%
\providecommand \bibitemStop [0]{}%
\providecommand \bibitemNoStop [0]{.\EOS\space}%
\providecommand \EOS [0]{\spacefactor3000\relax}%
\providecommand \BibitemShut  [1]{\csname bibitem#1\endcsname}%
\let\auto@bib@innerbib\@empty
%</preamble>
\bibitem [{\citenamefont {Vilenkin}(1980)}]{Vilenkin:1980fu}%
  \BibitemOpen
  \bibfield  {author} {\bibinfo {author} {\bibfnamefont {A.}~\bibnamefont
  {Vilenkin}},\ }\href {\doibase 10.1103/PhysRevD.22.3080} {\bibfield
  {journal} {\bibinfo  {journal} {Phys. Rev. D}\ }\textbf {\bibinfo {volume}
  {22}},\ \bibinfo {pages} {3080} (\bibinfo {year} {1980})}\BibitemShut
  {NoStop}%
\bibitem [{\citenamefont {Kharzeev}(2014)}]{Kharzeev:2013ffa}%
  \BibitemOpen
  \bibfield  {author} {\bibinfo {author} {\bibfnamefont {D.~E.}\ \bibnamefont
  {Kharzeev}},\ }\href {\doibase 10.1016/j.ppnp.2014.01.002} {\bibfield
  {journal} {\bibinfo  {journal} {Prog. Part. Nucl. Phys.}\ }\textbf {\bibinfo
  {volume} {75}},\ \bibinfo {pages} {133} (\bibinfo {year} {2014})},\ \Eprint
  {http://arxiv.org/abs/1312.3348} {arXiv:1312.3348 [hep-ph]} \BibitemShut
  {NoStop}%
\bibitem [{\citenamefont {Miransky}\ and\ \citenamefont
  {Shovkovy}(2015)}]{Miransky:2015ava}%
  \BibitemOpen
  \bibfield  {author} {\bibinfo {author} {\bibfnamefont {V.~A.}\ \bibnamefont
  {Miransky}}\ and\ \bibinfo {author} {\bibfnamefont {I.~A.}\ \bibnamefont
  {Shovkovy}},\ }\href {\doibase 10.1016/j.physrep.2015.02.003} {\bibfield
  {journal} {\bibinfo  {journal} {Phys. Rept.}\ }\textbf {\bibinfo {volume}
  {576}},\ \bibinfo {pages} {1} (\bibinfo {year} {2015})},\ \Eprint
  {http://arxiv.org/abs/1503.00732} {arXiv:1503.00732 [hep-ph]} \BibitemShut
  {NoStop}%
\bibitem [{\citenamefont {Kharzeev}(2022)}]{Kharzeev:2022ydx}%
  \BibitemOpen
  \bibfield  {author} {\bibinfo {author} {\bibfnamefont {D.~E.}\ \bibnamefont
  {Kharzeev}}\ }(\bibinfo {year} {2022})\ \Eprint
  {http://arxiv.org/abs/2204.10903} {arXiv:2204.10903 [hep-ph]} \BibitemShut
  {NoStop}%
\bibitem [{\citenamefont {Kharzeev}\ \emph {et~al.}(2008)\citenamefont
  {Kharzeev}, \citenamefont {McLerran},\ and\ \citenamefont
  {Warringa}}]{Kharzeev:2007jp}%
  \BibitemOpen
  \bibfield  {author} {\bibinfo {author} {\bibfnamefont {D.~E.}\ \bibnamefont
  {Kharzeev}}, \bibinfo {author} {\bibfnamefont {L.~D.}\ \bibnamefont
  {McLerran}}, \ and\ \bibinfo {author} {\bibfnamefont {H.~J.}\ \bibnamefont
  {Warringa}},\ }\href {\doibase 10.1016/j.nuclphysa.2008.02.298} {\bibfield
  {journal} {\bibinfo  {journal} {Nucl. Phys. A}\ }\textbf {\bibinfo {volume}
  {803}},\ \bibinfo {pages} {227} (\bibinfo {year} {2008})},\ \Eprint
  {http://arxiv.org/abs/0711.0950} {arXiv:0711.0950 [hep-ph]} \BibitemShut
  {NoStop}%
\bibitem [{\citenamefont {Fukushima}\ \emph {et~al.}(2008)\citenamefont
  {Fukushima}, \citenamefont {Kharzeev},\ and\ \citenamefont
  {Warringa}}]{Fukushima:2008xe}%
  \BibitemOpen
  \bibfield  {author} {\bibinfo {author} {\bibfnamefont {K.}~\bibnamefont
  {Fukushima}}, \bibinfo {author} {\bibfnamefont {D.~E.}\ \bibnamefont
  {Kharzeev}}, \ and\ \bibinfo {author} {\bibfnamefont {H.~J.}\ \bibnamefont
  {Warringa}},\ }\href {\doibase 10.1103/PhysRevD.78.074033} {\bibfield
  {journal} {\bibinfo  {journal} {Phys. Rev. D}\ }\textbf {\bibinfo {volume}
  {78}},\ \bibinfo {pages} {074033} (\bibinfo {year} {2008})},\ \Eprint
  {http://arxiv.org/abs/0808.3382} {arXiv:0808.3382 [hep-ph]} \BibitemShut
  {NoStop}%
\bibitem [{\citenamefont {Kharzeev}\ \emph {et~al.}(2013)\citenamefont
  {Kharzeev}, \citenamefont {Landsteiner}, \citenamefont {Schmitt},\ and\
  \citenamefont {Yee}}]{Kharzeev:2012ph}%
  \BibitemOpen
  \bibfield  {author} {\bibinfo {author} {\bibfnamefont {D.~E.}\ \bibnamefont
  {Kharzeev}}, \bibinfo {author} {\bibfnamefont {K.}~\bibnamefont
  {Landsteiner}}, \bibinfo {author} {\bibfnamefont {A.}~\bibnamefont
  {Schmitt}}, \ and\ \bibinfo {author} {\bibfnamefont {H.-U.}\ \bibnamefont
  {Yee}},\ }\href {\doibase 10.1007/978-3-642-37305-3_1} {\bibfield  {journal}
  {\bibinfo  {journal} {Lect. Notes Phys.}\ }\textbf {\bibinfo {volume}
  {871}},\ \bibinfo {pages} {1} (\bibinfo {year} {2013})},\ \Eprint
  {http://arxiv.org/abs/1211.6245} {arXiv:1211.6245 [hep-ph]} \BibitemShut
  {NoStop}%
\bibitem [{\citenamefont {Kharzeev}\ \emph {et~al.}(2016)\citenamefont
  {Kharzeev}, \citenamefont {Liao}, \citenamefont {Voloshin},\ and\
  \citenamefont {Wang}}]{Kharzeev:2015znc}%
  \BibitemOpen
  \bibfield  {author} {\bibinfo {author} {\bibfnamefont {D.~E.}\ \bibnamefont
  {Kharzeev}}, \bibinfo {author} {\bibfnamefont {J.}~\bibnamefont {Liao}},
  \bibinfo {author} {\bibfnamefont {S.~A.}\ \bibnamefont {Voloshin}}, \ and\
  \bibinfo {author} {\bibfnamefont {G.}~\bibnamefont {Wang}},\ }\href {\doibase
  10.1016/j.ppnp.2016.01.001} {\bibfield  {journal} {\bibinfo  {journal} {Prog.
  Part. Nucl. Phys.}\ }\textbf {\bibinfo {volume} {88}},\ \bibinfo {pages} {1}
  (\bibinfo {year} {2016})},\ \Eprint {http://arxiv.org/abs/1511.04050}
  {arXiv:1511.04050 [hep-ph]} \BibitemShut {NoStop}%
\bibitem [{\citenamefont {Koch}\ \emph {et~al.}(2017)\citenamefont {Koch},
  \citenamefont {Schlichting}, \citenamefont {Skokov}, \citenamefont
  {Sorensen}, \citenamefont {Thomas}, \citenamefont {Voloshin}, \citenamefont
  {Wang},\ and\ \citenamefont {Yee}}]{Koch:2016pzl}%
  \BibitemOpen
  \bibfield  {author} {\bibinfo {author} {\bibfnamefont {V.}~\bibnamefont
  {Koch}}, \bibinfo {author} {\bibfnamefont {S.}~\bibnamefont {Schlichting}},
  \bibinfo {author} {\bibfnamefont {V.}~\bibnamefont {Skokov}}, \bibinfo
  {author} {\bibfnamefont {P.}~\bibnamefont {Sorensen}}, \bibinfo {author}
  {\bibfnamefont {J.}~\bibnamefont {Thomas}}, \bibinfo {author} {\bibfnamefont
  {S.}~\bibnamefont {Voloshin}}, \bibinfo {author} {\bibfnamefont
  {G.}~\bibnamefont {Wang}}, \ and\ \bibinfo {author} {\bibfnamefont {H.-U.}\
  \bibnamefont {Yee}},\ }\href {\doibase 10.1088/1674-1137/41/7/072001}
  {\bibfield  {journal} {\bibinfo  {journal} {Chin. Phys. C}\ }\textbf
  {\bibinfo {volume} {41}},\ \bibinfo {pages} {072001} (\bibinfo {year}
  {2017})},\ \Eprint {http://arxiv.org/abs/1608.00982} {arXiv:1608.00982
  [nucl-th]} \BibitemShut {NoStop}%
\bibitem [{\citenamefont {Kharzeev}\ and\ \citenamefont
  {Liao}(2021)}]{Kharzeev:2020jxw}%
  \BibitemOpen
  \bibfield  {author} {\bibinfo {author} {\bibfnamefont {D.~E.}\ \bibnamefont
  {Kharzeev}}\ and\ \bibinfo {author} {\bibfnamefont {J.}~\bibnamefont
  {Liao}},\ }\href {\doibase 10.1038/s42254-020-00254-6} {\bibfield  {journal}
  {\bibinfo  {journal} {Nature Rev. Phys.}\ }\textbf {\bibinfo {volume} {3}},\
  \bibinfo {pages} {55} (\bibinfo {year} {2021})},\ \Eprint
  {http://arxiv.org/abs/2102.06623} {arXiv:2102.06623 [hep-ph]} \BibitemShut
  {NoStop}%
\bibitem [{\citenamefont {Son}\ and\ \citenamefont
  {Spivak}(2013)}]{Son:2012bg}%
  \BibitemOpen
  \bibfield  {author} {\bibinfo {author} {\bibfnamefont {D.~T.}\ \bibnamefont
  {Son}}\ and\ \bibinfo {author} {\bibfnamefont {B.~Z.}\ \bibnamefont
  {Spivak}},\ }\href {\doibase 10.1103/PhysRevB.88.104412} {\bibfield
  {journal} {\bibinfo  {journal} {Phys. Rev. B}\ }\textbf {\bibinfo {volume}
  {88}},\ \bibinfo {pages} {104412} (\bibinfo {year} {2013})},\ \Eprint
  {http://arxiv.org/abs/1206.1627} {arXiv:1206.1627 [cond-mat.mes-hall]}
  \BibitemShut {NoStop}%
\bibitem [{\citenamefont {Gorbar}\ \emph {et~al.}(2014)\citenamefont {Gorbar},
  \citenamefont {Miransky},\ and\ \citenamefont {Shovkovy}}]{Gorbar:2013dha}%
  \BibitemOpen
  \bibfield  {author} {\bibinfo {author} {\bibfnamefont {E.~V.}\ \bibnamefont
  {Gorbar}}, \bibinfo {author} {\bibfnamefont {V.~A.}\ \bibnamefont
  {Miransky}}, \ and\ \bibinfo {author} {\bibfnamefont {I.~A.}\ \bibnamefont
  {Shovkovy}},\ }\href {\doibase 10.1103/PhysRevB.89.085126} {\bibfield
  {journal} {\bibinfo  {journal} {Phys. Rev. B}\ }\textbf {\bibinfo {volume}
  {89}},\ \bibinfo {pages} {085126} (\bibinfo {year} {2014})},\ \Eprint
  {http://arxiv.org/abs/1312.0027} {arXiv:1312.0027 [cond-mat.mes-hall]}
  \BibitemShut {NoStop}%
\bibitem [{\citenamefont {Li}\ \emph {et~al.}(2016)\citenamefont {Li},
  \citenamefont {Kharzeev}, \citenamefont {Zhang}, \citenamefont {Huang},
  \citenamefont {Pletikosic}, \citenamefont {Fedorov}, \citenamefont {Zhong},
  \citenamefont {Schneeloch}, \citenamefont {Gu},\ and\ \citenamefont
  {Valla}}]{Li:2014bha}%
  \BibitemOpen
  \bibfield  {author} {\bibinfo {author} {\bibfnamefont {Q.}~\bibnamefont
  {Li}}, \bibinfo {author} {\bibfnamefont {D.~E.}\ \bibnamefont {Kharzeev}},
  \bibinfo {author} {\bibfnamefont {C.}~\bibnamefont {Zhang}}, \bibinfo
  {author} {\bibfnamefont {Y.}~\bibnamefont {Huang}}, \bibinfo {author}
  {\bibfnamefont {I.}~\bibnamefont {Pletikosic}}, \bibinfo {author}
  {\bibfnamefont {A.~V.}\ \bibnamefont {Fedorov}}, \bibinfo {author}
  {\bibfnamefont {R.~D.}\ \bibnamefont {Zhong}}, \bibinfo {author}
  {\bibfnamefont {J.~A.}\ \bibnamefont {Schneeloch}}, \bibinfo {author}
  {\bibfnamefont {G.~D.}\ \bibnamefont {Gu}}, \ and\ \bibinfo {author}
  {\bibfnamefont {T.}~\bibnamefont {Valla}},\ }\href {\doibase
  10.1038/nphys3648} {\bibfield  {journal} {\bibinfo  {journal} {Nature Phys.}\
  }\textbf {\bibinfo {volume} {12}},\ \bibinfo {pages} {550} (\bibinfo {year}
  {2016})},\ \Eprint {http://arxiv.org/abs/1412.6543} {arXiv:1412.6543
  [cond-mat.str-el]} \BibitemShut {NoStop}%
\bibitem [{\citenamefont {Cortijo}\ \emph {et~al.}(2016)\citenamefont
  {Cortijo}, \citenamefont {Kharzeev}, \citenamefont {Landsteiner},\ and\
  \citenamefont {Vozmediano}}]{Cortijo:2016wnf}%
  \BibitemOpen
  \bibfield  {author} {\bibinfo {author} {\bibfnamefont {A.}~\bibnamefont
  {Cortijo}}, \bibinfo {author} {\bibfnamefont {D.}~\bibnamefont {Kharzeev}},
  \bibinfo {author} {\bibfnamefont {K.}~\bibnamefont {Landsteiner}}, \ and\
  \bibinfo {author} {\bibfnamefont {M.~A.~H.}\ \bibnamefont {Vozmediano}},\
  }\href {\doibase 10.1103/PhysRevB.94.241405} {\bibfield  {journal} {\bibinfo
  {journal} {Phys. Rev. B}\ }\textbf {\bibinfo {volume} {94}},\ \bibinfo
  {pages} {241405} (\bibinfo {year} {2016})},\ \Eprint
  {http://arxiv.org/abs/1607.03491} {arXiv:1607.03491 [cond-mat.mes-hall]}
  \BibitemShut {NoStop}%
\bibitem [{\citenamefont {Kaushik}\ \emph {et~al.}(2019)\citenamefont
  {Kaushik}, \citenamefont {Kharzeev},\ and\ \citenamefont
  {Philip}}]{Kaushik:2018tjj}%
  \BibitemOpen
  \bibfield  {author} {\bibinfo {author} {\bibfnamefont {S.}~\bibnamefont
  {Kaushik}}, \bibinfo {author} {\bibfnamefont {D.~E.}\ \bibnamefont
  {Kharzeev}}, \ and\ \bibinfo {author} {\bibfnamefont {E.~J.}\ \bibnamefont
  {Philip}},\ }\href {\doibase 10.1103/PhysRevB.99.075150} {\bibfield
  {journal} {\bibinfo  {journal} {Phys. Rev. B}\ }\textbf {\bibinfo {volume}
  {99}},\ \bibinfo {pages} {075150} (\bibinfo {year} {2019})},\ \Eprint
  {http://arxiv.org/abs/1810.02399} {arXiv:1810.02399 [cond-mat.mes-hall]}
  \BibitemShut {NoStop}%
\bibitem [{\citenamefont {Sukhachov}\ \emph {et~al.}(2021)\citenamefont
  {Sukhachov}, \citenamefont {Gorbar},\ and\ \citenamefont
  {Shovkovy}}]{Sukhachov:2021fkh}%
  \BibitemOpen
  \bibfield  {author} {\bibinfo {author} {\bibfnamefont {P.~O.}\ \bibnamefont
  {Sukhachov}}, \bibinfo {author} {\bibfnamefont {E.~V.}\ \bibnamefont
  {Gorbar}}, \ and\ \bibinfo {author} {\bibfnamefont {I.~A.}\ \bibnamefont
  {Shovkovy}},\ }\href {\doibase 10.1103/PhysRevB.104.L121113} {\bibfield
  {journal} {\bibinfo  {journal} {Phys. Rev. B}\ }\textbf {\bibinfo {volume}
  {104}},\ \bibinfo {pages} {121113} (\bibinfo {year} {2021})},\ \Eprint
  {http://arxiv.org/abs/2103.15836} {arXiv:2103.15836 [cond-mat.mes-hall]}
  \BibitemShut {NoStop}%
\bibitem [{\citenamefont {Adler}(1969)}]{Adler:1969gk}%
  \BibitemOpen
  \bibfield  {author} {\bibinfo {author} {\bibfnamefont {S.~L.}\ \bibnamefont
  {Adler}},\ }\href {\doibase 10.1103/PhysRev.177.2426} {\bibfield  {journal}
  {\bibinfo  {journal} {Phys. Rev.}\ }\textbf {\bibinfo {volume} {177}},\
  \bibinfo {pages} {2426} (\bibinfo {year} {1969})}\BibitemShut {NoStop}%
\bibitem [{\citenamefont {Bell}\ and\ \citenamefont
  {Jackiw}(1969)}]{Bell:1969ts}%
  \BibitemOpen
  \bibfield  {author} {\bibinfo {author} {\bibfnamefont {J.~S.}\ \bibnamefont
  {Bell}}\ and\ \bibinfo {author} {\bibfnamefont {R.}~\bibnamefont {Jackiw}},\
  }\href {\doibase 10.1007/BF02823296} {\bibfield  {journal} {\bibinfo
  {journal} {Nuovo Cim. A}\ }\textbf {\bibinfo {volume} {60}},\ \bibinfo
  {pages} {47} (\bibinfo {year} {1969})}\BibitemShut {NoStop}%
\bibitem [{\citenamefont {Fujikawa}(1980)}]{Fujikawa:1980eg}%
  \BibitemOpen
  \bibfield  {author} {\bibinfo {author} {\bibfnamefont {K.}~\bibnamefont
  {Fujikawa}},\ }\href {\doibase 10.1103/PhysRevD.21.2848} {\bibfield
  {journal} {\bibinfo  {journal} {Phys. Rev. D}\ }\textbf {\bibinfo {volume}
  {21}},\ \bibinfo {pages} {2848} (\bibinfo {year} {1980})},\ \bibinfo {note}
  {[Erratum: Phys.Rev.D 22, 1499 (1980)]}\BibitemShut {NoStop}%
\bibitem [{\citenamefont {Hirono}\ \emph {et~al.}(2015)\citenamefont {Hirono},
  \citenamefont {Kharzeev},\ and\ \citenamefont {Yin}}]{Hirono:2015rla}%
  \BibitemOpen
  \bibfield  {author} {\bibinfo {author} {\bibfnamefont {Y.}~\bibnamefont
  {Hirono}}, \bibinfo {author} {\bibfnamefont {D.}~\bibnamefont {Kharzeev}}, \
  and\ \bibinfo {author} {\bibfnamefont {Y.}~\bibnamefont {Yin}},\ }\href
  {\doibase 10.1103/PhysRevD.92.125031} {\bibfield  {journal} {\bibinfo
  {journal} {Phys. Rev. D}\ }\textbf {\bibinfo {volume} {92}},\ \bibinfo
  {pages} {125031} (\bibinfo {year} {2015})},\ \Eprint
  {http://arxiv.org/abs/1509.07790} {arXiv:1509.07790 [hep-th]} \BibitemShut
  {NoStop}%
\bibitem [{\citenamefont {Gorbar}\ \emph {et~al.}(2016)\citenamefont {Gorbar},
  \citenamefont {Shovkovy}, \citenamefont {Vilchinskii}, \citenamefont
  {Rudenok}, \citenamefont {Boyarsky},\ and\ \citenamefont
  {Ruchayskiy}}]{Gorbar:2016qfh}%
  \BibitemOpen
  \bibfield  {author} {\bibinfo {author} {\bibfnamefont {E.~V.}\ \bibnamefont
  {Gorbar}}, \bibinfo {author} {\bibfnamefont {I.~A.}\ \bibnamefont
  {Shovkovy}}, \bibinfo {author} {\bibfnamefont {S.}~\bibnamefont
  {Vilchinskii}}, \bibinfo {author} {\bibfnamefont {I.}~\bibnamefont
  {Rudenok}}, \bibinfo {author} {\bibfnamefont {A.}~\bibnamefont {Boyarsky}}, \
  and\ \bibinfo {author} {\bibfnamefont {O.}~\bibnamefont {Ruchayskiy}},\
  }\href {\doibase 10.1103/PhysRevD.93.105028} {\bibfield  {journal} {\bibinfo
  {journal} {Phys. Rev. D}\ }\textbf {\bibinfo {volume} {93}},\ \bibinfo
  {pages} {105028} (\bibinfo {year} {2016})},\ \Eprint
  {http://arxiv.org/abs/1603.03442} {arXiv:1603.03442 [hep-th]} \BibitemShut
  {NoStop}%
\bibitem [{\citenamefont {Tuchin}(2018)}]{Tuchin:2017vwb}%
  \BibitemOpen
  \bibfield  {author} {\bibinfo {author} {\bibfnamefont {K.}~\bibnamefont
  {Tuchin}},\ }\href {\doibase 10.1016/j.nuclphysa.2017.09.015} {\bibfield
  {journal} {\bibinfo  {journal} {Nucl. Phys. A}\ }\textbf {\bibinfo {volume}
  {969}},\ \bibinfo {pages} {1} (\bibinfo {year} {2018})},\ \Eprint
  {http://arxiv.org/abs/1702.07329} {arXiv:1702.07329 [nucl-th]} \BibitemShut
  {NoStop}%
\bibitem [{\citenamefont {Mace}\ \emph {et~al.}(2020)\citenamefont {Mace},
  \citenamefont {Mueller}, \citenamefont {Schlichting},\ and\ \citenamefont
  {Sharma}}]{Mace:2019cqo}%
  \BibitemOpen
  \bibfield  {author} {\bibinfo {author} {\bibfnamefont {M.}~\bibnamefont
  {Mace}}, \bibinfo {author} {\bibfnamefont {N.}~\bibnamefont {Mueller}},
  \bibinfo {author} {\bibfnamefont {S.}~\bibnamefont {Schlichting}}, \ and\
  \bibinfo {author} {\bibfnamefont {S.}~\bibnamefont {Sharma}},\ }\href
  {\doibase 10.1103/PhysRevLett.124.191604} {\bibfield  {journal} {\bibinfo
  {journal} {Phys. Rev. Lett.}\ }\textbf {\bibinfo {volume} {124}},\ \bibinfo
  {pages} {191604} (\bibinfo {year} {2020})},\ \Eprint
  {http://arxiv.org/abs/1910.01654} {arXiv:1910.01654 [hep-ph]} \BibitemShut
  {NoStop}%
\bibitem [{\citenamefont {Joyce}\ and\ \citenamefont
  {Shaposhnikov}(1997)}]{Joyce:1997uy}%
  \BibitemOpen
  \bibfield  {author} {\bibinfo {author} {\bibfnamefont {M.}~\bibnamefont
  {Joyce}}\ and\ \bibinfo {author} {\bibfnamefont {M.~E.}\ \bibnamefont
  {Shaposhnikov}},\ }\href {\doibase 10.1103/PhysRevLett.79.1193} {\bibfield
  {journal} {\bibinfo  {journal} {Phys. Rev. Lett.}\ }\textbf {\bibinfo
  {volume} {79}},\ \bibinfo {pages} {1193} (\bibinfo {year} {1997})},\ \Eprint
  {http://arxiv.org/abs/astro-ph/9703005} {arXiv:astro-ph/9703005} \BibitemShut
  {NoStop}%
\bibitem [{\citenamefont {Gorbar}\ \emph
  {et~al.}(2018{\natexlab{a}})\citenamefont {Gorbar}, \citenamefont {Miransky},
  \citenamefont {Shovkovy},\ and\ \citenamefont {Sukhachov}}]{Gorbar:2017lnp}%
  \BibitemOpen
  \bibfield  {author} {\bibinfo {author} {\bibfnamefont {E.~V.}\ \bibnamefont
  {Gorbar}}, \bibinfo {author} {\bibfnamefont {V.~A.}\ \bibnamefont
  {Miransky}}, \bibinfo {author} {\bibfnamefont {I.~A.}\ \bibnamefont
  {Shovkovy}}, \ and\ \bibinfo {author} {\bibfnamefont {P.~O.}\ \bibnamefont
  {Sukhachov}},\ }\href {\doibase 10.1063/1.5037551} {\bibfield  {journal}
  {\bibinfo  {journal} {Low Temp. Phys.}\ }\textbf {\bibinfo {volume} {44}},\
  \bibinfo {pages} {487} (\bibinfo {year} {2018}{\natexlab{a}})},\ \Eprint
  {http://arxiv.org/abs/1712.08947} {arXiv:1712.08947 [cond-mat.mes-hall]}
  \BibitemShut {NoStop}%
\bibitem [{\citenamefont {Son}\ and\ \citenamefont
  {Yamamoto}(2012)}]{Son:2012wh}%
  \BibitemOpen
  \bibfield  {author} {\bibinfo {author} {\bibfnamefont {D.~T.}\ \bibnamefont
  {Son}}\ and\ \bibinfo {author} {\bibfnamefont {N.}~\bibnamefont {Yamamoto}},\
  }\href {\doibase 10.1103/PhysRevLett.109.181602} {\bibfield  {journal}
  {\bibinfo  {journal} {Phys. Rev. Lett.}\ }\textbf {\bibinfo {volume} {109}},\
  \bibinfo {pages} {181602} (\bibinfo {year} {2012})},\ \Eprint
  {http://arxiv.org/abs/1203.2697} {arXiv:1203.2697 [cond-mat.mes-hall]}
  \BibitemShut {NoStop}%
\bibitem [{\citenamefont {Manuel}\ and\ \citenamefont
  {Torres-Rincon}(2015)}]{Manuel:2015zpa}%
  \BibitemOpen
  \bibfield  {author} {\bibinfo {author} {\bibfnamefont {C.}~\bibnamefont
  {Manuel}}\ and\ \bibinfo {author} {\bibfnamefont {J.~M.}\ \bibnamefont
  {Torres-Rincon}},\ }\href {\doibase 10.1103/PhysRevD.92.074018} {\bibfield
  {journal} {\bibinfo  {journal} {Phys. Rev. D}\ }\textbf {\bibinfo {volume}
  {92}},\ \bibinfo {pages} {074018} (\bibinfo {year} {2015})},\ \Eprint
  {http://arxiv.org/abs/1501.07608} {arXiv:1501.07608 [hep-ph]} \BibitemShut
  {NoStop}%
\bibitem [{\citenamefont {Hattori}\ \emph {et~al.}(2019)\citenamefont
  {Hattori}, \citenamefont {Hirono}, \citenamefont {Yee},\ and\ \citenamefont
  {Yin}}]{Hattori:2017usa}%
  \BibitemOpen
  \bibfield  {author} {\bibinfo {author} {\bibfnamefont {K.}~\bibnamefont
  {Hattori}}, \bibinfo {author} {\bibfnamefont {Y.}~\bibnamefont {Hirono}},
  \bibinfo {author} {\bibfnamefont {H.-U.}\ \bibnamefont {Yee}}, \ and\
  \bibinfo {author} {\bibfnamefont {Y.}~\bibnamefont {Yin}},\ }\href {\doibase
  10.1103/PhysRevD.100.065023} {\bibfield  {journal} {\bibinfo  {journal}
  {Phys. Rev. D}\ }\textbf {\bibinfo {volume} {100}},\ \bibinfo {pages}
  {065023} (\bibinfo {year} {2019})},\ \Eprint
  {http://arxiv.org/abs/1711.08450} {arXiv:1711.08450 [hep-th]} \BibitemShut
  {NoStop}%
\bibitem [{\citenamefont {Gorbar}\ \emph
  {et~al.}(2018{\natexlab{b}})\citenamefont {Gorbar}, \citenamefont {Miransky},
  \citenamefont {Shovkovy},\ and\ \citenamefont {Sukhachov}}]{Gorbar:2017vph}%
  \BibitemOpen
  \bibfield  {author} {\bibinfo {author} {\bibfnamefont {E.~V.}\ \bibnamefont
  {Gorbar}}, \bibinfo {author} {\bibfnamefont {V.~A.}\ \bibnamefont
  {Miransky}}, \bibinfo {author} {\bibfnamefont {I.~A.}\ \bibnamefont
  {Shovkovy}}, \ and\ \bibinfo {author} {\bibfnamefont {P.~O.}\ \bibnamefont
  {Sukhachov}},\ }\href {\doibase 10.1103/PhysRevB.97.121105} {\bibfield
  {journal} {\bibinfo  {journal} {Phys. Rev. B}\ }\textbf {\bibinfo {volume}
  {97}},\ \bibinfo {pages} {121105} (\bibinfo {year} {2018}{\natexlab{b}})},\
  \Eprint {http://arxiv.org/abs/1712.01289} {arXiv:1712.01289
  [cond-mat.str-el]} \BibitemShut {NoStop}%
\bibitem [{\citenamefont {Buividovich}\ and\ \citenamefont
  {Ulybyshev}(2016)}]{Buividovich:2015jfa}%
  \BibitemOpen
  \bibfield  {author} {\bibinfo {author} {\bibfnamefont {P.~V.}\ \bibnamefont
  {Buividovich}}\ and\ \bibinfo {author} {\bibfnamefont {M.~V.}\ \bibnamefont
  {Ulybyshev}},\ }\href {\doibase 10.1103/PhysRevD.94.025009} {\bibfield
  {journal} {\bibinfo  {journal} {Phys. Rev. D}\ }\textbf {\bibinfo {volume}
  {94}},\ \bibinfo {pages} {025009} (\bibinfo {year} {2016})},\ \Eprint
  {http://arxiv.org/abs/1509.02076} {arXiv:1509.02076 [hep-th]} \BibitemShut
  {NoStop}%
\bibitem [{\citenamefont {Figueroa}\ \emph {et~al.}(2019)\citenamefont
  {Figueroa}, \citenamefont {Florio},\ and\ \citenamefont
  {Shaposhnikov}}]{Figueroa:2019jsi}%
  \BibitemOpen
  \bibfield  {author} {\bibinfo {author} {\bibfnamefont {D.~G.}\ \bibnamefont
  {Figueroa}}, \bibinfo {author} {\bibfnamefont {A.}~\bibnamefont {Florio}}, \
  and\ \bibinfo {author} {\bibfnamefont {M.}~\bibnamefont {Shaposhnikov}},\
  }\href {\doibase 10.1007/JHEP10(2019)142} {\bibfield  {journal} {\bibinfo
  {journal} {JHEP}\ }\textbf {\bibinfo {volume} {10}},\ \bibinfo {pages} {142}
  (\bibinfo {year} {2019})},\ \Eprint {http://arxiv.org/abs/1904.11892}
  {arXiv:1904.11892 [hep-th]} \BibitemShut {NoStop}%
\bibitem [{\citenamefont {Akamatsu}\ and\ \citenamefont
  {Yamamoto}(2013)}]{Akamatsu:2013pjd}%
  \BibitemOpen
  \bibfield  {author} {\bibinfo {author} {\bibfnamefont {Y.}~\bibnamefont
  {Akamatsu}}\ and\ \bibinfo {author} {\bibfnamefont {N.}~\bibnamefont
  {Yamamoto}},\ }\href {\doibase 10.1103/PhysRevLett.111.052002} {\bibfield
  {journal} {\bibinfo  {journal} {Phys. Rev. Lett.}\ }\textbf {\bibinfo
  {volume} {111}},\ \bibinfo {pages} {052002} (\bibinfo {year} {2013})},\
  \Eprint {http://arxiv.org/abs/1302.2125} {arXiv:1302.2125 [nucl-th]}
  \BibitemShut {NoStop}%
\bibitem [{\citenamefont {Rogachevskii}\ \emph {et~al.}(2017)\citenamefont
  {Rogachevskii}, \citenamefont {Ruchayskiy}, \citenamefont {Boyarsky},
  \citenamefont {Fr\"ohlich}, \citenamefont {Kleeorin}, \citenamefont
  {Brandenburg},\ and\ \citenamefont {Schober}}]{Rogachevskii:2017uyc}%
  \BibitemOpen
  \bibfield  {author} {\bibinfo {author} {\bibfnamefont {I.}~\bibnamefont
  {Rogachevskii}}, \bibinfo {author} {\bibfnamefont {O.}~\bibnamefont
  {Ruchayskiy}}, \bibinfo {author} {\bibfnamefont {A.}~\bibnamefont
  {Boyarsky}}, \bibinfo {author} {\bibfnamefont {J.}~\bibnamefont
  {Fr\"ohlich}}, \bibinfo {author} {\bibfnamefont {N.}~\bibnamefont
  {Kleeorin}}, \bibinfo {author} {\bibfnamefont {A.}~\bibnamefont
  {Brandenburg}}, \ and\ \bibinfo {author} {\bibfnamefont {J.}~\bibnamefont
  {Schober}},\ }\href {\doibase 10.3847/1538-4357/aa886b} {\bibfield  {journal}
  {\bibinfo  {journal} {Astrophys. J.}\ }\textbf {\bibinfo {volume} {846}},\
  \bibinfo {pages} {153} (\bibinfo {year} {2017})},\ \Eprint
  {http://arxiv.org/abs/1705.00378} {arXiv:1705.00378 [physics.plasm-ph]}
  \BibitemShut {NoStop}%
\bibitem [{\citenamefont {Mace}\ \emph {et~al.}(2016)\citenamefont {Mace},
  \citenamefont {Schlichting},\ and\ \citenamefont
  {Venugopalan}}]{Mace:2016svc}%
  \BibitemOpen
  \bibfield  {author} {\bibinfo {author} {\bibfnamefont {M.}~\bibnamefont
  {Mace}}, \bibinfo {author} {\bibfnamefont {S.}~\bibnamefont {Schlichting}}, \
  and\ \bibinfo {author} {\bibfnamefont {R.}~\bibnamefont {Venugopalan}},\
  }\href {\doibase 10.1103/PhysRevD.93.074036} {\bibfield  {journal} {\bibinfo
  {journal} {Phys. Rev. D}\ }\textbf {\bibinfo {volume} {93}},\ \bibinfo
  {pages} {074036} (\bibinfo {year} {2016})},\ \Eprint
  {http://arxiv.org/abs/1601.07342} {arXiv:1601.07342 [hep-ph]} \BibitemShut
  {NoStop}%
\bibitem [{\citenamefont {Lenz}(2005)}]{Lenz:2001me}%
  \BibitemOpen
  \bibfield  {author} {\bibinfo {author} {\bibfnamefont {F.}~\bibnamefont
  {Lenz}},\ }\href {\doibase 10.1007/978-3-540-31532-2_2} {\bibfield  {journal}
  {\bibinfo  {journal} {Lect. Notes Phys.}\ }\textbf {\bibinfo {volume}
  {659}},\ \bibinfo {pages} {7} (\bibinfo {year} {2005})},\ \Eprint
  {http://arxiv.org/abs/hep-th/0403286} {arXiv:hep-th/0403286} \BibitemShut
  {NoStop}%
\bibitem [{\citenamefont {Klinkhamer}\ and\ \citenamefont
  {Manton}(1984)}]{Klinkhamer:1984di}%
  \BibitemOpen
  \bibfield  {author} {\bibinfo {author} {\bibfnamefont {F.~R.}\ \bibnamefont
  {Klinkhamer}}\ and\ \bibinfo {author} {\bibfnamefont {N.~S.}\ \bibnamefont
  {Manton}},\ }\href {\doibase 10.1103/PhysRevD.30.2212} {\bibfield  {journal}
  {\bibinfo  {journal} {Phys. Rev. D}\ }\textbf {\bibinfo {volume} {30}},\
  \bibinfo {pages} {2212} (\bibinfo {year} {1984})}\BibitemShut {NoStop}%
\bibitem [{\citenamefont {Dashen}\ \emph {et~al.}(1974)\citenamefont {Dashen},
  \citenamefont {Hasslacher},\ and\ \citenamefont {Neveu}}]{Dashen:1974ck}%
  \BibitemOpen
  \bibfield  {author} {\bibinfo {author} {\bibfnamefont {R.~F.}\ \bibnamefont
  {Dashen}}, \bibinfo {author} {\bibfnamefont {B.}~\bibnamefont {Hasslacher}},
  \ and\ \bibinfo {author} {\bibfnamefont {A.}~\bibnamefont {Neveu}},\ }\href
  {\doibase 10.1103/PhysRevD.10.4138} {\bibfield  {journal} {\bibinfo
  {journal} {Phys. Rev. D}\ }\textbf {\bibinfo {volume} {10}},\ \bibinfo
  {pages} {4138} (\bibinfo {year} {1974})}\BibitemShut {NoStop}%
\bibitem [{\citenamefont {McLerran}\ \emph {et~al.}(1991)\citenamefont
  {McLerran}, \citenamefont {Mottola},\ and\ \citenamefont
  {Shaposhnikov}}]{McLerran:1990de}%
  \BibitemOpen
  \bibfield  {author} {\bibinfo {author} {\bibfnamefont {L.~D.}\ \bibnamefont
  {McLerran}}, \bibinfo {author} {\bibfnamefont {E.}~\bibnamefont {Mottola}}, \
  and\ \bibinfo {author} {\bibfnamefont {M.~E.}\ \bibnamefont {Shaposhnikov}},\
  }\href {\doibase 10.1103/PhysRevD.43.2027} {\bibfield  {journal} {\bibinfo
  {journal} {Phys. Rev. D}\ }\textbf {\bibinfo {volume} {43}},\ \bibinfo
  {pages} {2027} (\bibinfo {year} {1991})}\BibitemShut {NoStop}%
\bibitem [{\citenamefont {Cline}\ \emph {et~al.}(1993)\citenamefont {Cline},
  \citenamefont {Kainulainen},\ and\ \citenamefont {Olive}}]{Cline:1993vv}%
  \BibitemOpen
  \bibfield  {author} {\bibinfo {author} {\bibfnamefont {J.~M.}\ \bibnamefont
  {Cline}}, \bibinfo {author} {\bibfnamefont {K.}~\bibnamefont {Kainulainen}},
  \ and\ \bibinfo {author} {\bibfnamefont {K.~A.}\ \bibnamefont {Olive}},\
  }\href {\doibase 10.1103/PhysRevLett.71.2372} {\bibfield  {journal} {\bibinfo
   {journal} {Phys. Rev. Lett.}\ }\textbf {\bibinfo {volume} {71}},\ \bibinfo
  {pages} {2372} (\bibinfo {year} {1993})},\ \Eprint
  {http://arxiv.org/abs/hep-ph/9304321} {arXiv:hep-ph/9304321} \BibitemShut
  {NoStop}%
\bibitem [{\citenamefont {Kuzmin}\ \emph {et~al.}(1985)\citenamefont {Kuzmin},
  \citenamefont {Rubakov},\ and\ \citenamefont {Shaposhnikov}}]{Kuzmin:1985mm}%
  \BibitemOpen
  \bibfield  {author} {\bibinfo {author} {\bibfnamefont {V.~A.}\ \bibnamefont
  {Kuzmin}}, \bibinfo {author} {\bibfnamefont {V.~A.}\ \bibnamefont {Rubakov}},
  \ and\ \bibinfo {author} {\bibfnamefont {M.~E.}\ \bibnamefont
  {Shaposhnikov}},\ }\href {\doibase 10.1016/0370-2693(85)91028-7} {\bibfield
  {journal} {\bibinfo  {journal} {Phys. Lett. B}\ }\textbf {\bibinfo {volume}
  {155}},\ \bibinfo {pages} {36} (\bibinfo {year} {1985})}\BibitemShut
  {NoStop}%
\bibitem [{\citenamefont {Rubakov}\ and\ \citenamefont
  {Shaposhnikov}(1996)}]{Rubakov:1996vz}%
  \BibitemOpen
  \bibfield  {author} {\bibinfo {author} {\bibfnamefont {V.~A.}\ \bibnamefont
  {Rubakov}}\ and\ \bibinfo {author} {\bibfnamefont {M.~E.}\ \bibnamefont
  {Shaposhnikov}},\ }\href {\doibase 10.1070/PU1996v039n05ABEH000145}
  {\bibfield  {journal} {\bibinfo  {journal} {Usp. Fiz. Nauk}\ }\textbf
  {\bibinfo {volume} {166}},\ \bibinfo {pages} {493} (\bibinfo {year}
  {1996})},\ \Eprint {http://arxiv.org/abs/hep-ph/9603208}
  {arXiv:hep-ph/9603208} \BibitemShut {NoStop}%
\bibitem [{\citenamefont {Bodeker}\ and\ \citenamefont
  {Buchmuller}(2021)}]{Bodeker:2020ghk}%
  \BibitemOpen
  \bibfield  {author} {\bibinfo {author} {\bibfnamefont {D.}~\bibnamefont
  {Bodeker}}\ and\ \bibinfo {author} {\bibfnamefont {W.}~\bibnamefont
  {Buchmuller}},\ }\href {\doibase 10.1103/RevModPhys.93.035004} {\bibfield
  {journal} {\bibinfo  {journal} {Rev. Mod. Phys.}\ }\textbf {\bibinfo {volume}
  {93}},\ \bibinfo {pages} {035004} (\bibinfo {year} {2021})},\ \Eprint
  {http://arxiv.org/abs/2009.07294} {arXiv:2009.07294 [hep-ph]} \BibitemShut
  {NoStop}%
\bibitem [{\citenamefont {Bodeker}(1998)}]{Bodeker:1998hm}%
  \BibitemOpen
  \bibfield  {author} {\bibinfo {author} {\bibfnamefont {D.}~\bibnamefont
  {Bodeker}},\ }\href {\doibase 10.1016/S0370-2693(98)00279-2} {\bibfield
  {journal} {\bibinfo  {journal} {Phys. Lett. B}\ }\textbf {\bibinfo {volume}
  {426}},\ \bibinfo {pages} {351} (\bibinfo {year} {1998})},\ \Eprint
  {http://arxiv.org/abs/hep-ph/9801430} {arXiv:hep-ph/9801430} \BibitemShut
  {NoStop}%
\bibitem [{\citenamefont {Erdmenger}\ \emph {et~al.}(2009)\citenamefont
  {Erdmenger}, \citenamefont {Haack}, \citenamefont {Kaminski},\ and\
  \citenamefont {Yarom}}]{Erdmenger:2008rm}%
  \BibitemOpen
  \bibfield  {author} {\bibinfo {author} {\bibfnamefont {J.}~\bibnamefont
  {Erdmenger}}, \bibinfo {author} {\bibfnamefont {M.}~\bibnamefont {Haack}},
  \bibinfo {author} {\bibfnamefont {M.}~\bibnamefont {Kaminski}}, \ and\
  \bibinfo {author} {\bibfnamefont {A.}~\bibnamefont {Yarom}},\ }\href
  {\doibase 10.1088/1126-6708/2009/01/055} {\bibfield  {journal} {\bibinfo
  {journal} {JHEP}\ }\textbf {\bibinfo {volume} {01}},\ \bibinfo {pages} {055}
  (\bibinfo {year} {2009})},\ \Eprint {http://arxiv.org/abs/0809.2488}
  {arXiv:0809.2488 [hep-th]} \BibitemShut {NoStop}%
\bibitem [{\citenamefont {Son}\ and\ \citenamefont
  {Surowka}(2009)}]{Son:2009tf}%
  \BibitemOpen
  \bibfield  {author} {\bibinfo {author} {\bibfnamefont {D.~T.}\ \bibnamefont
  {Son}}\ and\ \bibinfo {author} {\bibfnamefont {P.}~\bibnamefont {Surowka}},\
  }\href {\doibase 10.1103/PhysRevLett.103.191601} {\bibfield  {journal}
  {\bibinfo  {journal} {Phys. Rev. Lett.}\ }\textbf {\bibinfo {volume} {103}},\
  \bibinfo {pages} {191601} (\bibinfo {year} {2009})},\ \Eprint
  {http://arxiv.org/abs/0906.5044} {arXiv:0906.5044 [hep-th]} \BibitemShut
  {NoStop}%
\bibitem [{\citenamefont {Loganayagam}(2011)}]{Loganayagam:2011mu}%
  \BibitemOpen
  \bibfield  {author} {\bibinfo {author} {\bibfnamefont {R.}~\bibnamefont
  {Loganayagam}},\ }\href@noop {} {\  (\bibinfo {year} {2011})},\ \Eprint
  {http://arxiv.org/abs/1106.0277} {arXiv:1106.0277 [hep-th]} \BibitemShut
  {NoStop}%
\bibitem [{\citenamefont {Kharzeev}\ and\ \citenamefont
  {Yee}(2011)}]{Kharzeev:2011ds}%
  \BibitemOpen
  \bibfield  {author} {\bibinfo {author} {\bibfnamefont {D.~E.}\ \bibnamefont
  {Kharzeev}}\ and\ \bibinfo {author} {\bibfnamefont {H.-U.}\ \bibnamefont
  {Yee}},\ }\href {\doibase 10.1103/PhysRevD.84.045025} {\bibfield  {journal}
  {\bibinfo  {journal} {Phys. Rev. D}\ }\textbf {\bibinfo {volume} {84}},\
  \bibinfo {pages} {045025} (\bibinfo {year} {2011})},\ \Eprint
  {http://arxiv.org/abs/1105.6360} {arXiv:1105.6360 [hep-th]} \BibitemShut
  {NoStop}%
\bibitem [{\citenamefont {Banerjee}\ \emph {et~al.}(2012)\citenamefont
  {Banerjee}, \citenamefont {Bhattacharya}, \citenamefont {Bhattacharyya},
  \citenamefont {Jain}, \citenamefont {Minwalla},\ and\ \citenamefont
  {Sharma}}]{Banerjee:2012iz}%
  \BibitemOpen
  \bibfield  {author} {\bibinfo {author} {\bibfnamefont {N.}~\bibnamefont
  {Banerjee}}, \bibinfo {author} {\bibfnamefont {J.}~\bibnamefont
  {Bhattacharya}}, \bibinfo {author} {\bibfnamefont {S.}~\bibnamefont
  {Bhattacharyya}}, \bibinfo {author} {\bibfnamefont {S.}~\bibnamefont {Jain}},
  \bibinfo {author} {\bibfnamefont {S.}~\bibnamefont {Minwalla}}, \ and\
  \bibinfo {author} {\bibfnamefont {T.}~\bibnamefont {Sharma}},\ }\href
  {\doibase 10.1007/JHEP09(2012)046} {\bibfield  {journal} {\bibinfo  {journal}
  {JHEP}\ }\textbf {\bibinfo {volume} {09}},\ \bibinfo {pages} {046} (\bibinfo
  {year} {2012})},\ \Eprint {http://arxiv.org/abs/1203.3544} {arXiv:1203.3544
  [hep-th]} \BibitemShut {NoStop}%
\bibitem [{\citenamefont {Hongo}\ \emph {et~al.}(2017)\citenamefont {Hongo},
  \citenamefont {Hirono},\ and\ \citenamefont {Hirano}}]{Hongo:2013cqa}%
  \BibitemOpen
  \bibfield  {author} {\bibinfo {author} {\bibfnamefont {M.}~\bibnamefont
  {Hongo}}, \bibinfo {author} {\bibfnamefont {Y.}~\bibnamefont {Hirono}}, \
  and\ \bibinfo {author} {\bibfnamefont {T.}~\bibnamefont {Hirano}},\ }\href
  {\doibase 10.1016/j.physletb.2017.10.028} {\bibfield  {journal} {\bibinfo
  {journal} {Phys. Lett. B}\ }\textbf {\bibinfo {volume} {775}},\ \bibinfo
  {pages} {266} (\bibinfo {year} {2017})},\ \Eprint
  {http://arxiv.org/abs/1309.2823} {arXiv:1309.2823 [nucl-th]} \BibitemShut
  {NoStop}%
\bibitem [{\citenamefont {Akamatsu}\ and\ \citenamefont
  {Yamamoto}(2014)}]{Akamatsu:2014yza}%
  \BibitemOpen
  \bibfield  {author} {\bibinfo {author} {\bibfnamefont {Y.}~\bibnamefont
  {Akamatsu}}\ and\ \bibinfo {author} {\bibfnamefont {N.}~\bibnamefont
  {Yamamoto}},\ }\href {\doibase 10.1103/PhysRevD.90.125031} {\bibfield
  {journal} {\bibinfo  {journal} {Phys. Rev. D}\ }\textbf {\bibinfo {volume}
  {90}},\ \bibinfo {pages} {125031} (\bibinfo {year} {2014})},\ \Eprint
  {http://arxiv.org/abs/1402.4174} {arXiv:1402.4174 [hep-th]} \BibitemShut
  {NoStop}%
\bibitem [{\citenamefont {Akamatsu}\ \emph {et~al.}(2016)\citenamefont
  {Akamatsu}, \citenamefont {Rothkopf},\ and\ \citenamefont
  {Yamamoto}}]{Akamatsu:2015kau}%
  \BibitemOpen
  \bibfield  {author} {\bibinfo {author} {\bibfnamefont {Y.}~\bibnamefont
  {Akamatsu}}, \bibinfo {author} {\bibfnamefont {A.}~\bibnamefont {Rothkopf}},
  \ and\ \bibinfo {author} {\bibfnamefont {N.}~\bibnamefont {Yamamoto}},\
  }\href {\doibase 10.1007/JHEP03(2016)210} {\bibfield  {journal} {\bibinfo
  {journal} {JHEP}\ }\textbf {\bibinfo {volume} {03}},\ \bibinfo {pages} {210}
  (\bibinfo {year} {2016})},\ \Eprint {http://arxiv.org/abs/1512.02374}
  {arXiv:1512.02374 [hep-ph]} \BibitemShut {NoStop}%
\bibitem [{\citenamefont {Mueller}\ and\ \citenamefont
  {Venugopalan}(2017)}]{Mueller:2017arw}%
  \BibitemOpen
  \bibfield  {author} {\bibinfo {author} {\bibfnamefont {N.}~\bibnamefont
  {Mueller}}\ and\ \bibinfo {author} {\bibfnamefont {R.}~\bibnamefont
  {Venugopalan}},\ }\href {\doibase 10.1103/PhysRevD.96.016023} {\bibfield
  {journal} {\bibinfo  {journal} {Phys. Rev. D}\ }\textbf {\bibinfo {volume}
  {96}},\ \bibinfo {pages} {016023} (\bibinfo {year} {2017})},\ \Eprint
  {http://arxiv.org/abs/1702.01233} {arXiv:1702.01233 [hep-ph]} \BibitemShut
  {NoStop}%
\bibitem [{\citenamefont {Polkovnikov}(2010)}]{Polkovnikov:2009ys}%
  \BibitemOpen
  \bibfield  {author} {\bibinfo {author} {\bibfnamefont {A.}~\bibnamefont
  {Polkovnikov}},\ }\href {\doibase 10.1016/j.aop.2010.02.006} {\bibfield
  {journal} {\bibinfo  {journal} {Annals Phys.}\ }\textbf {\bibinfo {volume}
  {325}},\ \bibinfo {pages} {1790} (\bibinfo {year} {2010})},\ \Eprint
  {http://arxiv.org/abs/0905.3384} {arXiv:0905.3384 [cond-mat.stat-mech]}
  \BibitemShut {NoStop}%
\bibitem [{\citenamefont {Berges}\ \emph {et~al.}(2012)\citenamefont {Berges},
  \citenamefont {Schlichting},\ and\ \citenamefont {Sexty}}]{Berges:2012ev}%
  \BibitemOpen
  \bibfield  {author} {\bibinfo {author} {\bibfnamefont {J.}~\bibnamefont
  {Berges}}, \bibinfo {author} {\bibfnamefont {S.}~\bibnamefont {Schlichting}},
  \ and\ \bibinfo {author} {\bibfnamefont {D.}~\bibnamefont {Sexty}},\ }\href
  {\doibase 10.1103/PhysRevD.86.074006} {\bibfield  {journal} {\bibinfo
  {journal} {Phys. Rev. D}\ }\textbf {\bibinfo {volume} {86}},\ \bibinfo
  {pages} {074006} (\bibinfo {year} {2012})},\ \Eprint
  {http://arxiv.org/abs/1203.4646} {arXiv:1203.4646 [hep-ph]} \BibitemShut
  {NoStop}%
\bibitem [{\citenamefont {Kurkela}\ and\ \citenamefont
  {Moore}(2012)}]{Kurkela:2012hp}%
  \BibitemOpen
  \bibfield  {author} {\bibinfo {author} {\bibfnamefont {A.}~\bibnamefont
  {Kurkela}}\ and\ \bibinfo {author} {\bibfnamefont {G.~D.}\ \bibnamefont
  {Moore}},\ }\href {\doibase 10.1103/PhysRevD.86.056008} {\bibfield  {journal}
  {\bibinfo  {journal} {Phys. Rev. D}\ }\textbf {\bibinfo {volume} {86}},\
  \bibinfo {pages} {056008} (\bibinfo {year} {2012})},\ \Eprint
  {http://arxiv.org/abs/1207.1663} {arXiv:1207.1663 [hep-ph]} \BibitemShut
  {NoStop}%
\bibitem [{\citenamefont {Jeon}(2014)}]{Jeon:2013zga}%
  \BibitemOpen
  \bibfield  {author} {\bibinfo {author} {\bibfnamefont {S.}~\bibnamefont
  {Jeon}},\ }\href {\doibase 10.1016/j.aop.2013.09.019} {\bibfield  {journal}
  {\bibinfo  {journal} {Annals Phys.}\ }\textbf {\bibinfo {volume} {340}},\
  \bibinfo {pages} {119} (\bibinfo {year} {2014})},\ \Eprint
  {http://arxiv.org/abs/1308.0263} {arXiv:1308.0263 [hep-th]} \BibitemShut
  {NoStop}%
\bibitem [{\citenamefont {Berges}(2015)}]{Berges:2015kfa}%
  \BibitemOpen
  \bibfield  {author} {\bibinfo {author} {\bibfnamefont {J.}~\bibnamefont
  {Berges}},\ }\href@noop {} {\  (\bibinfo {year} {2015})},\ \Eprint
  {http://arxiv.org/abs/1503.02907} {arXiv:1503.02907 [hep-ph]} \BibitemShut
  {NoStop}%
\bibitem [{\citenamefont {Aarts}\ and\ \citenamefont
  {Smit}(1999)}]{Aarts:1998td}%
  \BibitemOpen
  \bibfield  {author} {\bibinfo {author} {\bibfnamefont {G.}~\bibnamefont
  {Aarts}}\ and\ \bibinfo {author} {\bibfnamefont {J.}~\bibnamefont {Smit}},\
  }\href {\doibase 10.1016/S0550-3213(99)00320-X} {\bibfield  {journal}
  {\bibinfo  {journal} {Nucl. Phys. B}\ }\textbf {\bibinfo {volume} {555}},\
  \bibinfo {pages} {355} (\bibinfo {year} {1999})},\ \Eprint
  {http://arxiv.org/abs/hep-ph/9812413} {arXiv:hep-ph/9812413} \BibitemShut
  {NoStop}%
\bibitem [{\citenamefont {Berges}\ \emph {et~al.}(2011)\citenamefont {Berges},
  \citenamefont {Gelfand},\ and\ \citenamefont {Pruschke}}]{Berges:2010zv}%
  \BibitemOpen
  \bibfield  {author} {\bibinfo {author} {\bibfnamefont {J.}~\bibnamefont
  {Berges}}, \bibinfo {author} {\bibfnamefont {D.}~\bibnamefont {Gelfand}}, \
  and\ \bibinfo {author} {\bibfnamefont {J.}~\bibnamefont {Pruschke}},\ }\href
  {\doibase 10.1103/PhysRevLett.107.061301} {\bibfield  {journal} {\bibinfo
  {journal} {Phys. Rev. Lett.}\ }\textbf {\bibinfo {volume} {107}},\ \bibinfo
  {pages} {061301} (\bibinfo {year} {2011})},\ \Eprint
  {http://arxiv.org/abs/1012.4632} {arXiv:1012.4632 [hep-ph]} \BibitemShut
  {NoStop}%
\bibitem [{\citenamefont {Kasper}\ \emph {et~al.}(2014)\citenamefont {Kasper},
  \citenamefont {Hebenstreit},\ and\ \citenamefont {Berges}}]{Kasper:2014uaa}%
  \BibitemOpen
  \bibfield  {author} {\bibinfo {author} {\bibfnamefont {V.}~\bibnamefont
  {Kasper}}, \bibinfo {author} {\bibfnamefont {F.}~\bibnamefont {Hebenstreit}},
  \ and\ \bibinfo {author} {\bibfnamefont {J.}~\bibnamefont {Berges}},\ }\href
  {\doibase 10.1103/PhysRevD.90.025016} {\bibfield  {journal} {\bibinfo
  {journal} {Phys. Rev. D}\ }\textbf {\bibinfo {volume} {90}},\ \bibinfo
  {pages} {025016} (\bibinfo {year} {2014})},\ \Eprint
  {http://arxiv.org/abs/1403.4849} {arXiv:1403.4849 [hep-ph]} \BibitemShut
  {NoStop}%
\bibitem [{\citenamefont {Mueller}\ \emph {et~al.}(2016)\citenamefont
  {Mueller}, \citenamefont {Hebenstreit},\ and\ \citenamefont
  {Berges}}]{Mueller:2016aao}%
  \BibitemOpen
  \bibfield  {author} {\bibinfo {author} {\bibfnamefont {N.}~\bibnamefont
  {Mueller}}, \bibinfo {author} {\bibfnamefont {F.}~\bibnamefont
  {Hebenstreit}}, \ and\ \bibinfo {author} {\bibfnamefont {J.}~\bibnamefont
  {Berges}},\ }\href {\doibase 10.1103/PhysRevLett.117.061601} {\bibfield
  {journal} {\bibinfo  {journal} {Phys. Rev. Lett.}\ }\textbf {\bibinfo
  {volume} {117}},\ \bibinfo {pages} {061601} (\bibinfo {year} {2016})},\
  \Eprint {http://arxiv.org/abs/1605.01413} {arXiv:1605.01413 [hep-ph]}
  \BibitemShut {NoStop}%
\bibitem [{\citenamefont {M\"uller}\ \emph {et~al.}(2016)\citenamefont
  {M\"uller}, \citenamefont {Schlichting},\ and\ \citenamefont
  {Sharma}}]{Muller:2016jod}%
  \BibitemOpen
  \bibfield  {author} {\bibinfo {author} {\bibfnamefont {N.}~\bibnamefont
  {M\"uller}}, \bibinfo {author} {\bibfnamefont {S.}~\bibnamefont
  {Schlichting}}, \ and\ \bibinfo {author} {\bibfnamefont {S.}~\bibnamefont
  {Sharma}},\ }\href {\doibase 10.1103/PhysRevLett.117.142301} {\bibfield
  {journal} {\bibinfo  {journal} {Phys. Rev. Lett.}\ }\textbf {\bibinfo
  {volume} {117}},\ \bibinfo {pages} {142301} (\bibinfo {year} {2016})},\
  \Eprint {http://arxiv.org/abs/1606.00342} {arXiv:1606.00342 [hep-ph]}
  \BibitemShut {NoStop}%
\bibitem [{\citenamefont {Mace}\ \emph {et~al.}(2017)\citenamefont {Mace},
  \citenamefont {Mueller}, \citenamefont {Schlichting},\ and\ \citenamefont
  {Sharma}}]{Mace:2016shq}%
  \BibitemOpen
  \bibfield  {author} {\bibinfo {author} {\bibfnamefont {M.}~\bibnamefont
  {Mace}}, \bibinfo {author} {\bibfnamefont {N.}~\bibnamefont {Mueller}},
  \bibinfo {author} {\bibfnamefont {S.}~\bibnamefont {Schlichting}}, \ and\
  \bibinfo {author} {\bibfnamefont {S.}~\bibnamefont {Sharma}},\ }\href
  {\doibase 10.1103/PhysRevD.95.036023} {\bibfield  {journal} {\bibinfo
  {journal} {Phys. Rev. D}\ }\textbf {\bibinfo {volume} {95}},\ \bibinfo
  {pages} {036023} (\bibinfo {year} {2017})},\ \Eprint
  {http://arxiv.org/abs/1612.02477} {arXiv:1612.02477 [hep-lat]} \BibitemShut
  {NoStop}%
\bibitem [{\citenamefont {Kogut}\ and\ \citenamefont
  {Susskind}(1975)}]{Kogut:1974ag}%
  \BibitemOpen
  \bibfield  {author} {\bibinfo {author} {\bibfnamefont {J.~B.}\ \bibnamefont
  {Kogut}}\ and\ \bibinfo {author} {\bibfnamefont {L.}~\bibnamefont
  {Susskind}},\ }\href {\doibase 10.1103/PhysRevD.11.395} {\bibfield  {journal}
  {\bibinfo  {journal} {Phys. Rev. D}\ }\textbf {\bibinfo {volume} {11}},\
  \bibinfo {pages} {395} (\bibinfo {year} {1975})}\BibitemShut {NoStop}%
\bibitem [{\citenamefont {Aarts}\ and\ \citenamefont
  {Berges}(2002)}]{Aarts:2001yn}%
  \BibitemOpen
  \bibfield  {author} {\bibinfo {author} {\bibfnamefont {G.}~\bibnamefont
  {Aarts}}\ and\ \bibinfo {author} {\bibfnamefont {J.}~\bibnamefont {Berges}},\
  }\href {\doibase 10.1103/PhysRevLett.88.041603} {\bibfield  {journal}
  {\bibinfo  {journal} {Phys. Rev. Lett.}\ }\textbf {\bibinfo {volume} {88}},\
  \bibinfo {pages} {041603} (\bibinfo {year} {2002})},\ \Eprint
  {http://arxiv.org/abs/hep-ph/0107129} {arXiv:hep-ph/0107129} \BibitemShut
  {NoStop}%
\bibitem [{\citenamefont {Berges}\ \emph {et~al.}(2014)\citenamefont {Berges},
  \citenamefont {Boguslavski}, \citenamefont {Schlichting},\ and\ \citenamefont
  {Venugopalan}}]{Berges:2013fga}%
  \BibitemOpen
  \bibfield  {author} {\bibinfo {author} {\bibfnamefont {J.}~\bibnamefont
  {Berges}}, \bibinfo {author} {\bibfnamefont {K.}~\bibnamefont {Boguslavski}},
  \bibinfo {author} {\bibfnamefont {S.}~\bibnamefont {Schlichting}}, \ and\
  \bibinfo {author} {\bibfnamefont {R.}~\bibnamefont {Venugopalan}},\ }\href
  {\doibase 10.1103/PhysRevD.89.114007} {\bibfield  {journal} {\bibinfo
  {journal} {Phys. Rev. D}\ }\textbf {\bibinfo {volume} {89}},\ \bibinfo
  {pages} {114007} (\bibinfo {year} {2014})},\ \Eprint
  {http://arxiv.org/abs/1311.3005} {arXiv:1311.3005 [hep-ph]} \BibitemShut
  {NoStop}%
\bibitem [{\citenamefont {Berges}\ \emph {et~al.}(2008)\citenamefont {Berges},
  \citenamefont {Scheffler},\ and\ \citenamefont {Sexty}}]{Berges:2007re}%
  \BibitemOpen
  \bibfield  {author} {\bibinfo {author} {\bibfnamefont {J.}~\bibnamefont
  {Berges}}, \bibinfo {author} {\bibfnamefont {S.}~\bibnamefont {Scheffler}}, \
  and\ \bibinfo {author} {\bibfnamefont {D.}~\bibnamefont {Sexty}},\ }\href
  {\doibase 10.1103/PhysRevD.77.034504} {\bibfield  {journal} {\bibinfo
  {journal} {Phys. Rev. D}\ }\textbf {\bibinfo {volume} {77}},\ \bibinfo
  {pages} {034504} (\bibinfo {year} {2008})},\ \Eprint
  {http://arxiv.org/abs/0712.3514} {arXiv:0712.3514 [hep-ph]} \BibitemShut
  {NoStop}%
\bibitem [{\citenamefont {Kurkela}\ and\ \citenamefont
  {Moore}(2011)}]{Kurkela:2011ti}%
  \BibitemOpen
  \bibfield  {author} {\bibinfo {author} {\bibfnamefont {A.}~\bibnamefont
  {Kurkela}}\ and\ \bibinfo {author} {\bibfnamefont {G.~D.}\ \bibnamefont
  {Moore}},\ }\href {\doibase 10.1007/JHEP12(2011)044} {\bibfield  {journal}
  {\bibinfo  {journal} {JHEP}\ }\textbf {\bibinfo {volume} {12}},\ \bibinfo
  {pages} {044} (\bibinfo {year} {2011})},\ \Eprint
  {http://arxiv.org/abs/1107.5050} {arXiv:1107.5050 [hep-ph]} \BibitemShut
  {NoStop}%
\bibitem [{\citenamefont {Karsten}\ and\ \citenamefont
  {Smit}(1981)}]{Karsten:1980wd}%
  \BibitemOpen
  \bibfield  {author} {\bibinfo {author} {\bibfnamefont {L.~H.}\ \bibnamefont
  {Karsten}}\ and\ \bibinfo {author} {\bibfnamefont {J.}~\bibnamefont {Smit}},\
  }\href {\doibase 10.1016/0550-3213(81)90549-6} {\bibfield  {journal}
  {\bibinfo  {journal} {Nucl. Phys. B}\ }\textbf {\bibinfo {volume} {183}},\
  \bibinfo {pages} {103} (\bibinfo {year} {1981})}\BibitemShut {NoStop}%
\bibitem [{\citenamefont {Moore}(1999)}]{Moore:1998swa}%
  \BibitemOpen
  \bibfield  {author} {\bibinfo {author} {\bibfnamefont {G.~D.}\ \bibnamefont
  {Moore}},\ }\href {\doibase 10.1103/PhysRevD.59.014503} {\bibfield  {journal}
  {\bibinfo  {journal} {Phys. Rev. D}\ }\textbf {\bibinfo {volume} {59}},\
  \bibinfo {pages} {014503} (\bibinfo {year} {1999})},\ \Eprint
  {http://arxiv.org/abs/hep-ph/9805264} {arXiv:hep-ph/9805264} \BibitemShut
  {NoStop}%
\bibitem [{\citenamefont {Barroso~Mancha}\ and\ \citenamefont
  {Moore}(2023)}]{BarrosoMancha:2022mbj}%
  \BibitemOpen
  \bibfield  {author} {\bibinfo {author} {\bibfnamefont {M.}~\bibnamefont
  {Barroso~Mancha}}\ and\ \bibinfo {author} {\bibfnamefont {G.~D.}\
  \bibnamefont {Moore}},\ }\href {\doibase 10.1007/JHEP01(2023)155} {\bibfield
  {journal} {\bibinfo  {journal} {JHEP}\ }\textbf {\bibinfo {volume} {01}},\
  \bibinfo {pages} {155} (\bibinfo {year} {2023})},\ \Eprint
  {http://arxiv.org/abs/2210.05507} {arXiv:2210.05507 [hep-lat]} \BibitemShut
  {NoStop}%
\bibitem [{\citenamefont {Altenkort}\ \emph {et~al.}(2021)\citenamefont
  {Altenkort}, \citenamefont {Eller}, \citenamefont {Kaczmarek}, \citenamefont
  {Mazur}, \citenamefont {Moore},\ and\ \citenamefont
  {Shu}}]{Altenkort:2020axj}%
  \BibitemOpen
  \bibfield  {author} {\bibinfo {author} {\bibfnamefont {L.}~\bibnamefont
  {Altenkort}}, \bibinfo {author} {\bibfnamefont {A.~M.}\ \bibnamefont
  {Eller}}, \bibinfo {author} {\bibfnamefont {O.}~\bibnamefont {Kaczmarek}},
  \bibinfo {author} {\bibfnamefont {L.}~\bibnamefont {Mazur}}, \bibinfo
  {author} {\bibfnamefont {G.~D.}\ \bibnamefont {Moore}}, \ and\ \bibinfo
  {author} {\bibfnamefont {H.-T.}\ \bibnamefont {Shu}},\ }\href {\doibase
  10.1103/PhysRevD.103.114513} {\bibfield  {journal} {\bibinfo  {journal}
  {Phys. Rev. D}\ }\textbf {\bibinfo {volume} {103}},\ \bibinfo {pages}
  {114513} (\bibinfo {year} {2021})},\ \Eprint
  {http://arxiv.org/abs/2012.08279} {arXiv:2012.08279 [hep-lat]} \BibitemShut
  {NoStop}%
\bibitem [{\citenamefont {Moore}(1996)}]{Moore:1996wn}%
  \BibitemOpen
  \bibfield  {author} {\bibinfo {author} {\bibfnamefont {G.~D.}\ \bibnamefont
  {Moore}},\ }\href {\doibase 10.1016/S0550-3213(96)00497-X} {\bibfield
  {journal} {\bibinfo  {journal} {Nucl. Phys. B}\ }\textbf {\bibinfo {volume}
  {480}},\ \bibinfo {pages} {689} (\bibinfo {year} {1996})},\ \Eprint
  {http://arxiv.org/abs/hep-lat/9605001} {arXiv:hep-lat/9605001} \BibitemShut
  {NoStop}%
\end{thebibliography}%



\end{document}